\documentclass[12pt]{article}

\usepackage{scicite}
\usepackage{graphicx}

\usepackage{times}
\usepackage{amssymb}

\newenvironment{sciabstract}{%
\begin{quote} \bf}
{\end{quote}}

\title{Pendell\"{o}sung Interferometry Probes the Neutron Charge Radius, Lattice Dynamics, and Fifth Forces}

\author
{Benjamin Heacock,$^{1,2,3,\ast}$ Takuhiro Fujiie,$^{4,5}$ Robert W. Haun,$^{6,7}$ \\
Albert Henins,$^{1}$ Katsuya Hirota,$^{4,8}$ Takuya Hosobata,$^{5}$ Michael G. Huber,$^1$ \\ Masaaki Kitaguchi,$^{4,9}$ 
Dmitry A. Pushin,$^{10,11}$ Hirohiko Shimizu,$^4$ \\ Masahiro Takeda,$^5$ Robert Valdillez,$^{2,3}$
Yutaka Yamagata,$^5$ Albert R. Young$^{2,3}$\\
\\
\normalsize{$^{1}$National Institute of Standards and Technology, Gaithersburg, MD 20899, USA}\\
\normalsize{$^{2}$Department of Physics, North Carolina State University, Raleigh, NC 27695, USA}\\
\normalsize{$^{3}$Triangle Universities Nuclear Laboratory, Durham, NC 27708, USA}\\
\normalsize{$^{4}$Department of Physics, Nagoya University, Nagoya 464-8602, Japan}\\
\normalsize{$^{5}$RIKEN Center for Advanced Photonics, RIKEN, Hirosawa 2-1,}\\ 
\normalsize{Wako, Saitama 351-0198, Japan}\\
\normalsize{$^{6}$Institute for Physical Science and Technology, University of Maryland,}\\
\normalsize{College Park, MD 20742, USA} \\
\normalsize{$^7$Department of Physics and Engineering Physics, Tulane University,}\\ 
\normalsize{New Orleans, LA 70118, USA}\\
\normalsize{$^{8}$High Energy Accelerator Research Organization , Tsukuba 305-0801, Japan}\\
\normalsize{$^{9}$Kobayashi-Maskawa Institute, Nagoya University, Nagoya 464-8602, Japan}\\
\normalsize{$^{10}$Institute for Quantum Computing, University of Waterloo, Waterloo, ON, Canada, N2L3G1}\\
\normalsize{$^{11}$Department of Physics and Astronomy, University of Waterloo,}\\ \normalsize{Waterloo, ON, Canada, N2L3G1}\\
\\
\normalsize{$^\ast$To whom correspondence should be addressed; E-mail:  benjamin.heacock@nist.gov.}
}

\date{}

\begin{document} 

\baselineskip24pt


\maketitle


\begin{sciabstract}
\boldmath
Structure factors describe how incident radiation is scattered from materials such as silicon and germanium and characterize the physical interaction between the material and scattered particles. We use neutron pendell\"{o}sung interferometry to make precision measurements of the (220) and (400) neutron-silicon structure factors, and achieve a factor of four improvement in the (111) structure factor uncertainty. These data provide measurements of the silicon Debye-Waller factor at room temperature and the mean square neutron charge radius  $\langle r_n^2 \rangle = -0.1101 \pm 0.0089 \, \mathrm{fm}^2$. Combined with existing measurements of the Debye-Waller factor and charge radius, the measured structure factors also improve constraints on the strength of a Yukawa-modification to gravity by an order of magnitude over the 20~pm to 10~nm length scale range.  
\unboldmath
\end{sciabstract}



\noindent
Neutrons, electrons, and x-rays all exhibit pendell\"{o}sung interference upon Bragg diffraction in the Laue geometry, where the Bragg planes are perpendicular to the entrance and exit faces of the crystal (Fig.~1A).   The phenomenon, characterized by oscillations of diffracted intensity as a function of probe wavelength, crystal thickness, and lattice potential, was predicted by Ewald \cite{ewald1916begrundung} and first observed in electron diffraction from MgO \cite{heidenreich1942electron}; it was later demonstrated using x-rays \cite{kato1959study} and neutrons \cite{sippel1965pendellosungs,shull1968observation}.  X-ray pendell\"{o}sung studies in particular \cite{kato1959study,hattori1965absolute,tanemura1972absolute, wada1977intensity,deutsch1985new} have proved an invaluable tool for measuring the electron density in silicon \cite{willis1969lattice,aldred1973electron,spackman1986electron,cummings1988redetermination} and provide data which can be compared to {\it ab initio} lattice dynamical models \cite{erba2013accurate}.  Neutron pendell\"{o}sung measurements yield information complementary to that obtained with other probes; however, achieving high-precision has proven challenging.


A neutron wave propagating along the symmetry planes of a Bragg-diffracting crystal forms standing waves with integer periodicity to the Bragg planes. The neutron’s kinematic momentum then depends on the incident neutron energy and the overlap of the standing wave with the crystalline potential. The detuning of the incident neutron wave into energetically degenerate “fast” and “slow” modes creates a spatial beating period along the Bragg planes between the diffracted and transmitted intensities which may be resolved as pendell\"{o}sung oscillations. The absolute phase shift of the  pendell\"{o}sung interference fringes is determined by the material's structure factor, which is unique to the radiation species and Miller indices $(hkl)$ of the Bragg planes.  For the neutron case, structure factors can be described in terms of the thermally-averaged single atom coherent elastic scattering amplitudes $b(Q) = 2 m e^{-W} \widetilde{V}(Q) / \hbar^2 $, where $\hbar$ is the reduced Planck constant, $m$ is the neutron mass, and $\widetilde{V}(Q)$ is the Fourier transform of the neutron-atomic potential.  The Debye-Waller factor ($e^{-W}$, DWF) accounts for the thermal motion of the atoms in the lattice, where $W =  B Q^2/(16 \pi^2)$ and $B$ is determined by the mean-square thermal atomic displacement $\langle u^2 \rangle = B / (8 \pi^2 )$. The momentum transfer magnitude is $Q_{hkl} = (2 \pi / a) \sqrt{h^2 + l^2 + k^2}$, where $a$ is the lattice constant for a material with a cubic unit cell. 


The main contribution to $b(Q)$ is the coherent nuclear scattering length, which is $Q$-independent. The leading-order $Q$-dependent contribution to $b(Q)$ (Fig.~1F) is the DWF, exhibiting a relative impact of a few percent.  A determination of $B$ can provide a benchmark for lattice dynamical models.  The next-to-leading-order contribution to the $Q$-dependence of $b(Q)$ arises from the neutron's spherically-symmetric charge distribution (which is a result of its three-quark composite structure) interacting with the very large interatomic electric fields of $\sim 10^8 \, \mathrm{V} \, \mathrm{cm}^{-1}$ \cite{fedorov2010measurement}.  This interaction is described by the mean square neutron charge radius $\langle r_n^2 \rangle$, where $- \langle r_n^2 \rangle / 6$ is the slope of the neutron electric form factor $G_E^n(Q^2)$ with respect to $Q^2$ at zero momentum transfer.  Determinations of $\langle r_n^2 \rangle $ along with electron scattering data can be used to study the neutron's internal charge distribution \cite{gentile2011neutron, atac2021charge, atac2021measurement} or as a parameter in chiral effective field theory (EFT) studies of light nuclei \cite{epelbaum2020high}.  As proposed by Ref~\cite{sparenberg2002neutron}, a pendell\"{o}sung inteferometry determination of $\langle r_n^2 \rangle$ can weigh in on the slight tension between individual neutron scattering experiments making up the Particle Data Group (PDG) recommended value  of $\langle r_n^2 \rangle_\mathrm{PDG} = -0.1161(22)$~fm$^2$ \cite{PDG2018}.

Aside from the expected $B$ and $\langle r_n^2 \rangle$ contributions, $b(Q)$ is sensitive to interactions beyond the standard model of particle physics, often referred to as BSM interactions.  BSM interactions on the atomic length scale would result in a unique $Q$-dependence \cite{greene2007neutron}.  
Yet-to-be-detected ``fifth" forces arise from several BSM theories seeking to explain mysteries of modern physics, including the incompatibility of general relativity and quantum mechanics \cite{arkani1999phenomenology, adelberger2003tests, mostepanenko2020state}, dark matter and/or dark energy \cite{zee2004dark, fayet2007u, fichet2018quantum, brax2018bounding, brax2019warped}, and the smallness of neutrino masses \cite{arkani2002new}.  The large variety of BSM theories has motivated a multi-disciplinary experimental effort to constrain yet-undiscovered physics over the entirety of physical observation scales (thirty-five orders of magnitude in length scale) from collider experiments at the shortest length scales and highest energies, to astrophysical observations at the largest length scales and lowest energies \cite{murata2015review}.

Here we tabulate the measured $b(Q_{hkl})$ for silicon and use these data to (1) make a high-precision determination of $B$ for silicon at 295.5~K, (2) measure the neutron charge radius, and (3) place constraints on the strength of a BSM Yukawa-modification of gravity over the 20~pm to 10~nm length scale range.  

\begin{figure*}[htb!]
    \centering
    \includegraphics[width = 0.95\textwidth]{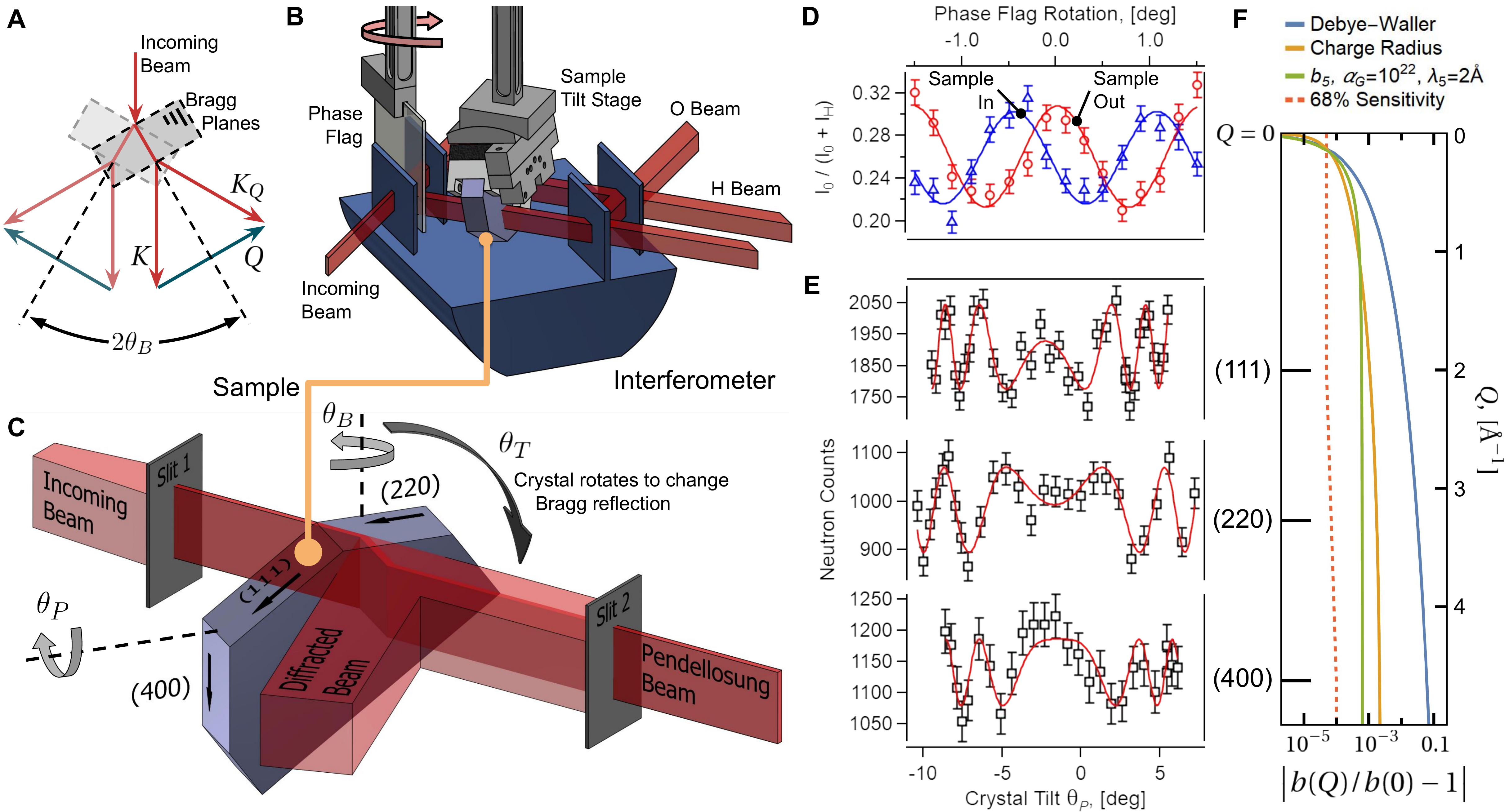}
    \linespread{2}
    \caption{\textbf{Pendell\"{o}sung interference in silicon.} \textbf{(A)} Bragg diffraction in the Laue geometry can be measured when the crystal is rotated by $\pm \theta_B$ relative to the incoming monochromatic beam. Here, $Q$ is the momentum transfer, and $K$ and $K_Q$ are the forward-diffracted and diffracted beams, respectively. \textbf{(B)} The quantity $b(0) D$ is measured using an interferometer. \textbf{(C)} $b(Q) D$ is measured using pendell\"{o}sung oscillations in the forward-diffracted beam. Note that the (400) arrow points in the (004) direction; we indicate Miller indices with the largest value first to emphasize that $b(Q)$ depends on the magnitude of $Q$.  \textbf{(D,E)} Typical interferograms with lines of best fit are shown for the interferometer and pendell\"{o}sung setups, respectively.  Error bars are from Poisson counting statistics.  \textbf{(F)} The theoretical relative contributions of each $Q$-dependent term in $b(Q)$; note that the Debye-Waller term is negative, whereas the other two terms are positive.  The data are sensitive (68 \%-level) to features larger than the dashed line.}
    \label{fig:fig1}
\end{figure*}

Pendell\"{o}sung interference may be resolved in either the diffracted or forward-diffracted beams ($K_Q$ and $K$ in Fig.~\ref{fig:fig1}A, respectively) by rotating a crystal slab precisely about the axis perpendicular to the diffraction plane (wavelength method \cite{shull1968observation}) or perpendicular to the Bragg planes (crystal thickness method \cite{somenkov1978observation}, used here), creating an interferogram (Fig.~1E).  The interferogram is then fit to a functional form (Eq.~S1) to determine the phase of the oscillations, which is proportional to $b(Q)$, the neutron wavelength $\lambda$, and the thickness of the crystal slab $D$.  We also made forward-scattering ($Q=0$) measurements for each sample orientation using a perfect-crystal neutron interferometer, a device in which a monolithic crystal connects Bragg-diffracting ``blades'' that protrude from a common base (Fig.~1B, see also \cite{rauch2015neutron}).  The phase shift between the two interferometer paths is modulated by rotating a fused silica phase flag, creating an interferogram (Fig~1D).  The phase flag interferograms are likewise fit to a functional form (Eq.~S4) and the phase is extracted \cite{rauch2015neutron}.  The forward-scattering phase shift from the sample is then the difference in the fitted phase between the sample-in and sample-out of the beam (Fig.~1D) and is proportional to $b(0)$, $\lambda$, and $D$.

The extraction of $b(Q) D$ from the measured phase shift was enabled by an in-situ determination of $\lambda$ in which pendell\"{o}sung interferograms were obtained with the crystal rotated by both positive and negative Bragg angles $\pm \theta_B$, as is shown in Fig.~1A.  The average phase shift is then quadratic in the slight difference in wavelength between the two interferograms and small enough to be neglected.  The average wavelength was computed using $\lambda = 4 \pi \sin \theta_B / Q_{hkl}$, the silicon lattice constant, and $2 \theta_B$ as given by an angular encoder embedded in the crystal's rotational positioning stage.  A similar approach for measuring wavelength was used for the forward scattering measurements, with the interferometer and accompanying optics rotated by $\pm \theta_B$ of the interferometer.



Prior measurements of $b(Q_{111})$ were subject to strain fields in the crystal slab, observed as a one-directional phase shift with unique wavelength-dependence \cite{shull1972spherical}.  However, the reported strain gradients were much larger than what is observed in neutron interferometers generally \cite{heacock2017neutron,heacock2018increased} and specifically in a neutron interferometer that utilized a very similar machining and post-fabrication process \cite{heacock2019measurement} to the pendell\"{o}sung sample used here.  Reducing strain gradients invariably comes at the cost of increased variation in $D$, but a strain-relieving acid etch now typical for neutron interferometers was not performed in the prior measurement of $b(Q_{111})$ \cite{shull1972spherical}.  
To separate $b(Q)$ from $D$, we normalized the pendell\"{o}sung phase shift with the forward scattering phase shift measured over the same relevant crystal volume for each Bragg reflection.   To this end, flats were cut on the pendell\"{o}sung sample so that it could fit between the interferometer optical components with the neutron beam still illuminating the required sample area.  The ratio of the two measurements forms $b(Q)/ b(0)$, naturally isolating the $Q$-dependence of the scattering amplitude while also eliminating the need to measure $D$ by other means.  This normalization relaxes the requisite crystal flatness, thereby enabling measurements of $b(Q)/b(0)$ using an acid-etched, strain-free crystal.



Both pendell\"{o}sung and forward scattering measurements were performed at the neutron interferometer and optics facility auxiliary beamline (NIOFa) at the National Institute of Standards and Technology (NIST) Center for Neutron Research (NCNR).  The NIOFa provides both 2.2~\AA~and 4.4~\AA~monochromatic ($\Delta \lambda / \lambda \sim 0.5$~\%) neutrons \cite{shahi2016new}.  The (220) and (400) pendell\"{o}sung interferograms were measured using 2.2~\AA, and the (111) pendell\"{o}sung interferograms and interferometer forward scattering measurements used 4.4~\AA. The resulting ratios $b(Q) / b(0)$ for the (111), (220), and (400) Bragg reflections are reported in Table~1.  

The theoretical shape versus $Q$ of the contributions to $b(Q)/b(0)$ from the DWF and $\langle r_n^2 \rangle$ are shown in Fig.~1F.  To extract these two parameters, the three measured values of $b(Q)/b(0)$ were fit to

\begin{equation}
    {b(Q) \over b(0)} = e^{-W} \left ( 1 - Z {b_{ne} \over b(0)} \right ) + f_e(hkl) {b_{ne} \over b(0)}
    \label{Eqn:bq}
\end{equation}

\noindent
by minimizing the $\chi^2$ sum of weighted residuals, where $Z$ is the crystal's atomic number and $B$ and $b_{ne}$ are treated as fit parameters.  The room temperature x-ray silicon form factors $f_e(hkl)$ in units of elementary charge are given by x-ray scattering measurements \cite{cummings1988redetermination}. The neutron-electron scattering length $b_{ne}$ is related to the charge radius by $b_{ne} = \langle r_n^2 \rangle m_n / (3 m_e a_0) = \langle r^2_n \rangle  / (86.34 \; \mathrm{fm} )$, with the electron mass $m_e$ and Bohr radius $a_0$. The fit function's sensitivity to the forward scattering length $b(0)$ \cite{ioffe1998precision} and $f_e(hkl)$  \cite{cummings1988redetermination} is small compared to experimental uncertainties.

If $B$ and $\langle r_n^2 \rangle$ are treated as free parameters, then our results are $B = 0.4761(17)$~\AA$^2$ and $\langle r_n^2 \rangle = -0.1101(89)$~fm$^2$ with a correlation coefficient of -0.94; the $\chi^2$-surface of $B$ and $\langle r^2_n \rangle$ versus prior determinations of both parameters is shown in Fig.~2.  Detailed comparisons are made in \cite{Supp}, but of particular interest is the lattice dynamical result from fitting neutron inelastic scattering data to a Born-von K\'{a}rm\'{a}n (BvK) model $B_\mathrm{BvK} = 0.4725(17)$~\AA$^2$ when scaled to 295.5~K \cite{flensburg1999lattice} and the world average $\langle r_n^2 \rangle _\mathrm{avg} = -0.1137(13) \, \mathrm{fm}^2$ from prior neutron scattering experiments, the computation of which is described in \cite{Supp}.  The confidence region formed by $B_\mathrm{BvK}$ and $\langle r_n^2 \rangle_\mathrm{avg}$ shows slight tension with our results.  

In regards to $\langle r_n^2 \rangle$, most prior determinations come from epithermal neutron total transmission through lead or bismuth, with potentially-correlated systematic uncertainties from solid state and nuclear resonance effects \cite{kopecky1997neutron}.  The only other type of experiment contributing to $\langle r_n^2 \rangle _\mathrm{PDG} $ and $\langle r_n^2 \rangle_\mathrm{avg}$ are scattering asymmetry measurements in noble gasses \cite{krohn1966measurement, krohn1973reconsideration}.  Pendell\"{o}sung interferometry constitutes the only determination of $\langle r_n^2 \rangle $ using cold neutrons and contains entirely-different systematic uncertainties compared to these previous methods \cite{sparenberg2002neutron,sparenberg2003neutron}.  Furthermore, a recent determination of $\langle r_n^2 \rangle_\mathrm{EFT} = -0.106^{+0.007}_{-0.005} \, \mathrm{fm}^2$ from chiral EFT combined with measurements of the hydrogen-deuterium isotope shift \cite{filin2020extraction} is consistent with our results and shifted in the same direction relative to $\langle r_n^2 \rangle_\mathrm{PDG}$, with $\langle r_n^2 \rangle_\mathrm{EFT}$ showing a $1.8 \sigma$ difference with $\langle r_n^2 \rangle_\mathrm{PDG}$ and a less-concerning $1.4 \sigma$ difference with $\langle r_n^2 \rangle_\mathrm{avg}$.  

It may be possible to decrease uncertainties in the extracted $\langle r_n^2\rangle$ by constraining $B$. However, the tendency for x-ray structure factor determinations of $B$ to increasingly deviate from theory for higher-order structure factors \cite{deutsch1989thermal,erba2013accurate}, as well as the disagreement between $B$ as determined by x-rays versus neutrons, suggests a breakdown in the rigid atom approximation \cite{Supp}.  On the other hand, neutron inelastic scattering data, as well as fully anharmonic lattice dynamical models, have shown that harmonic approximations used by the BvK and other models are insufficient for silicon's strong anharmonicity \cite{hellman2013temperature,kim2015phonon,kim2018nuclear}.  Furthermore, the BvK model was unable to predict the measured frequencies on the phonon dispersion curves for relative uncertainties less than 0.5~\% \cite{flensburg1999lattice}, suggesting an inadequacy of the BvK model at this level of precision, in which case the quoted relative uncertainty for $B_\mathrm{BvK}$ of 0.34~\% may be too small.  Whether semi-empirically fitting the interatomic force constants from other lattice dynamical models would produce better fits to neutron inelastic scattering data, and whether the resulting determination of $B$ in turn agrees with our results, could provide a means for evaluating different thermodynamic lattice models.  Because the reduced $\chi^2$ for the BvK model fit was $\chi^2_\mathrm{red} = 33.5$ (193-21 degrees of freedom) before the expansion of any experimental uncertainties \cite{flensburg1999lattice}, a particularly-successful lattice model may be able to reduce uncertainties in $B$ by up to a factor of $\sqrt{\chi^2_\mathrm{red}} = 5.8$ relative to $B_\mathrm{BvK}$.  If the resulting 0.06~\% measurement of $B$ were applied as a constraint to our evaluation of $\langle r_n^2 \rangle$, then the uncertainty thereof would be reduced to $\lesssim 0.004$~fm$^2$.    

A future test of lattice dynamical models using pendell\"{o}sung interferometry may be achieved from the sensitivity of odd structure factors to three-phonon terms via the anharmonic DWF (aDWF).  Of the reflections in Table~1, only the (111) reflection is affected by the aDWF and only at the $3 \times 10^{-5}$ level \cite{Supp}.  However, the contribution grows like $h \times k \times l$ \cite{flensburg1999lattice}, and a measurement of the (333) structure factor with a relative precision similar to Table~1 is capable of providing a determination of the average cubic displacement along the silicon bond direction $\langle u_{111}^3 \rangle$ at the 5~\% level.  The near room temperature (288~K) $\langle u_{111}^3 \rangle$ has only been measured with 10~\% relative uncertainty via the ``forbidden" (222) structure factor, which itself deviated from the expected $T^2$ temperature dependence extrapolated from higher temperatures ($> 650 $~K) by a factor of two \cite{roberto1974diffraction}, suggesting that quantum zero-point anharmonic vibrations may contribute meaningfully to $\langle u_{111}^3 \rangle$ at room temperature.  Pendell\"{o}sung interferometry is unique in its capability to be extended to cryogenic temperatures to measure $\langle u_{111}^3 \rangle$ in the regime where the anharmonic zero-point vibrations are the most relevant.  This is of particular interest for silicon, given how the anharmonic force constants determine silicon's electrical properties and anomalous thermal expansion at low temperatures \cite{kim2018nuclear}, providing motivation to measure higher-order odd $b(Q_{hkl})$ at a variety of temperatures.

A BSM ``fifth" fundamental force or other BSM physics adds an additional term

\begin{equation}
    \delta \left ( \frac{ b(Q)}{b(0)} \right ) = e^{-W}{b_5(Q) - b_5(0) \over b(0)}
    \label{eqn:b5Q}
\end{equation}

\noindent
to Eq.~\ref{Eqn:bq}, given a BSM contribution to the neutron-atomic potential $b_5(Q) = 2 m \widetilde{V}_5(Q) / \hbar^2$.  
The method for using neutron structure factors to constrain BSM physics is general:  A $\chi^2$ sum of weighted residuals for the measured structure factors, as well as $B$, the aDWF, and $ \langle r_n^2 \rangle $, is minimized with respect to BSM model parameters, given a functional form of $b_5(Q)$.  The weights and values assigned to the $B$, $\langle r_n^2 \rangle$, and the aDWF residuals are to be based on an estimation of how the BSM physics would bias the experiments from which the residuals are derived.  The tabulated structure factors can thus continue to constrain future BSM theories as they are developed, with improved constraints as the number of measured structure factors increases.

A common parameterization for constraining BSM physics is a Yukawa-modification to gravity \cite{adelberger2003tests}, leading to $\delta [ b(Q) / b(0) ]e^W = \alpha_G  ( 3.9 \times 10^{-27} \, \mathrm{\AA}^{-2} \, \lambda_5^2 ) Q^2 \lambda_5^2 / (1 + Q^2 \lambda_5^2 )$ for the case of silicon in Eq.~\ref{eqn:b5Q} \cite{greene2007neutron}; this expression depends on an interaction strength relative to gravity $\alpha_G$ and length scale $\lambda_5$.  The coupling to nucleon number $g_S$ of an undiscovered massive scalar with particle mass $m_5 = \hbar / (c \lambda_5)$, where $c$ is the speed of light, proposed by a number of BSM theories is constrained by the same limits \cite{adelberger2003tests, nesvizhevsky2008neutron, mostepanenko2020state}.  The constraints on the strength of an undiscovered Yukawa-potential over the relevant range of $\lambda_5$ from this work compared to previous experiments is shown in Fig.~3.  Our results constitute an order of magnitude improvement over nearly three decades in $\lambda_5$.  Previous limits constrain $\langle r_n^2 \rangle $ according to the PDG value and uncertainty \cite{kamiya2015constraints,haddock2018search}, except for a higher-energy experiment that uses an expanded uncertainty \cite{pokotilovski2006constraints}, 
where a systematic shift in the measured $\langle r_n^2 \rangle$ from an unaccounted-for Yukawa-modification to gravity could be relevant for $\lambda_5 \lesssim 10$~pm.  In keeping with other previously-published constraints, we likewise set the $\langle r_n^2 \rangle$ residual according to the PDG value and uncertainty when computing the limits in Fig.~3. 

In addition to measuring $b(Q)$ for higher-order reflections in silicon, measuring multiple $b(Q)$ from germanium, for which precision measurements only exist for the (111) reflection \cite{shull1973neutron}, is expected to improve constraints on $\alpha_G$ by a factor five, regardless of whether $\langle r_n^2 \rangle $ is constrained because of the differing $M/Z$ for germanium and silicon.  Germanium measurements would also reduce the uncertainty in our measured $\langle r_n^2 \rangle$ to $\sim 0.005$~fm$^2$ with $B$ from both atomic species unconstrained, in which case pendell\"{o}sung interferometry could weigh-in more strongly on the $\langle r_n^2 \rangle $ landscape.  If determinations of $B$ for silicon and germanium can be achieved at the 0.1~\%-level using neutron inelastic scattering and lattice dynamical models, then an achievable uncertainty for $\langle r_n^2 \rangle$ from pendell\"{o}sung interferometry of $0.002$~fm$^2$ would be comparable to the uncertainty of $\langle r_n^2 \rangle_\mathrm{PDG}$.  Ultimately, whereas x-ray structure factors in silicon as large as $(12 \, 12 \, 0)$ and (880) have been measured at room and liquid nitrogen temperatures, respectively  \cite{spackman1986electron,cummings1988redetermination}, an expanding set of tabulated neutron structure factors for silicon, germanium, and other crystals at multiple temperatures would provide increasingly-sensitive measures of material-specific thermal displacement parameters and the neutron charge radius, while continuing to improve BSM constraints.




\bibliographystyle{Science}



\section*{Acknowledgments}

We thank the NIST Center for Neutron Research for providing the neutron facilities used in this work.  \textbf{Funding:} This work was supported by The National Institute of Standards and Technology, the US Department of Energy (DOE) under Grant No. 89243019SSC000025 and Grant No. DE-FG02-97ER41042, the National Science Foundation (NSF) under Grant No. PHY-1307426, the Natural Sciences and Engineering Council of Canada (NSERC) Discovery program, and the Canada First Research Excellence Fund (CFREF). \textbf{Author contributions:} BH, MGH, HS, and ARY conceived the project. BH, TF, RWH, MGH, DAP, and RV collected neutron data under the supervision of MGH. AH performed the sample crystallographic alignment and initial fabrication. TF, TH, MT, and YY machined and characterized the sample surfaces under the supervision of YY. BH, MGH, and DAP chemically etched the sample. MGH, KH, MK, HS, YY, and ARY were responsible for overall supervision. BH performed the main analysis and wrote the manuscript with input from all authors. \textbf{Data and materials availability:} The datasets and analysis code are available through Zenodo \cite{Data}.


\section*{Supplementary materials}
Materials and Methods\\
Supplementary Text\\
Figs. S1 to S5\\
Tables S1 and S2\\




\clearpage

\clearpage

\begin{figure*}[htb!]
    \centering
    \includegraphics[width = \textwidth]{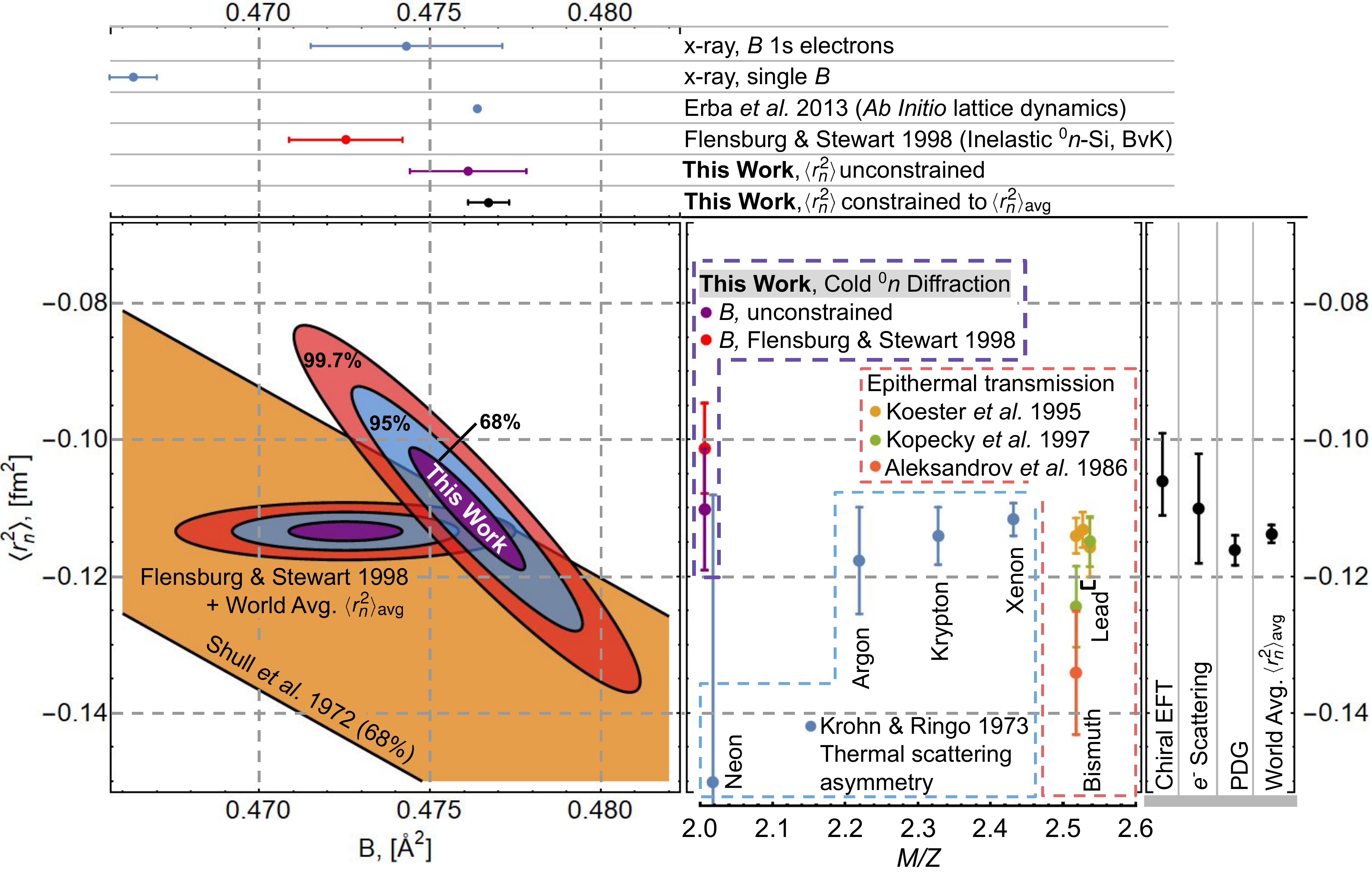}
    \linespread{2}
    \caption{\textbf{ \boldmath Confidence regions over $\langle r_n^2 \rangle$ and $B$. \unboldmath} Data have been rescaled to 295.5~K and compared to previous work. Our data are consistent with the previous (111) silicon measurement Shull et al. 1972 \cite{shull1972spherical}, but have four times the precision.  The Erba et al. 2013 \cite{erba2013accurate} determination of $B$ is a theoretically-computed value which does not have an estimated uncertainty. Other previous work includes Flensburg and Stewart 1998 \cite{flensburg1999lattice}, Krohn and Ringo 1973 \cite{krohn1973reconsideration}, Koester et al. 1995 \cite{koester1995neutron}, Kopecky et al. 1997 \cite{kopecky1997neutron}, Aleksandrov et al. 1986 \cite{aleksandrov1986neutron}, Chiral EFT \cite{epelbaum2020high}, $e^-$ Scattering \cite{atac2021measurement}, PDG \cite{PDG2018}, and World Avg. \cite{Supp}.  See \cite{Supp} for a detailed discussion.}
    \label{fig:Results}
\end{figure*}

\clearpage

\begin{figure*}[htb!]
    \centering
    \includegraphics[width = \textwidth]{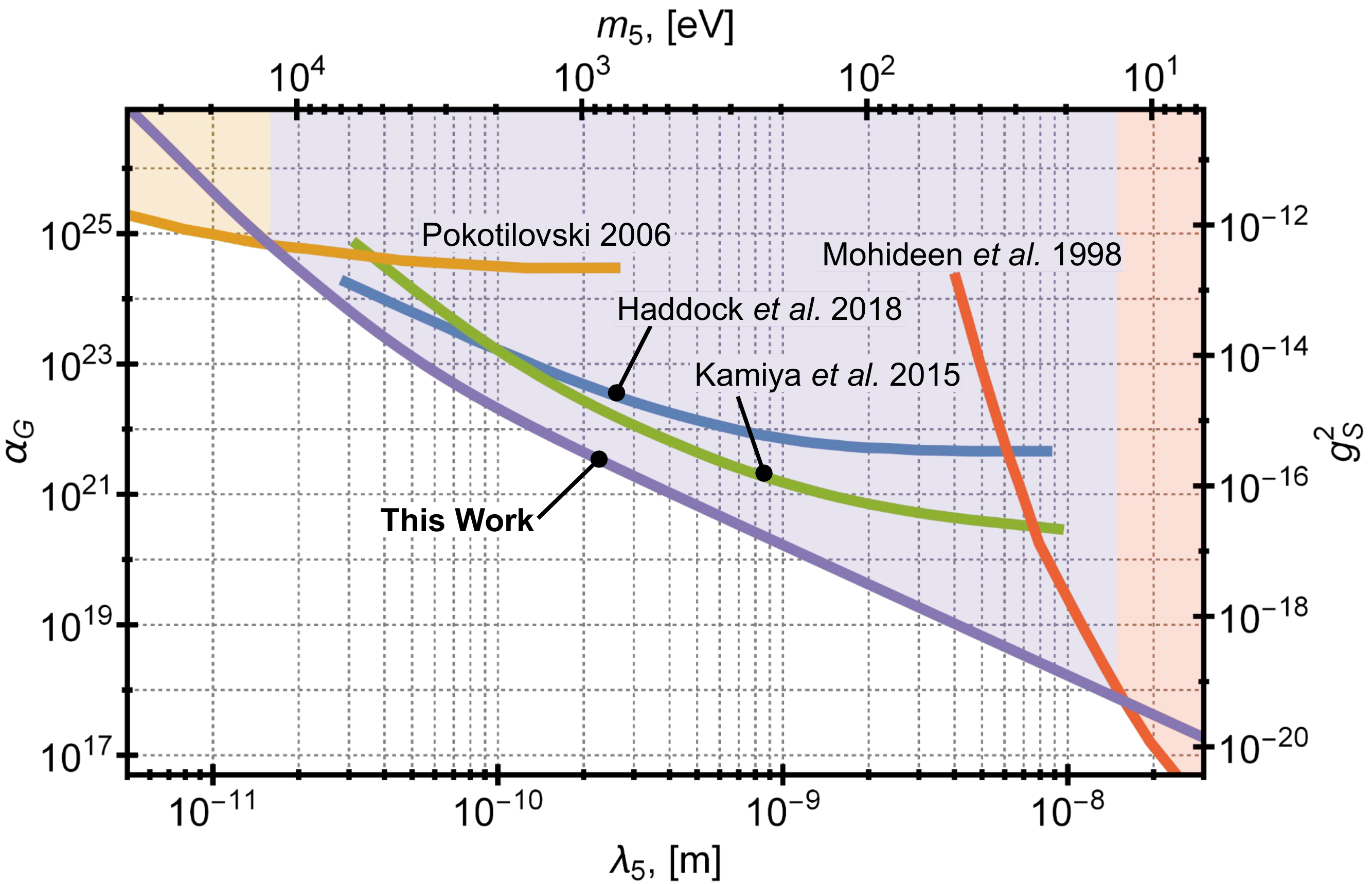}
    \linespread{2}
    \caption{\textbf{ \boldmath Limits (95~\% confidence) on the strength of a Yukawa-modification to gravity $\alpha_G$ compared to previous experiments as a function of the force's range $\lambda_5$. \unboldmath }  The same limits constrain the coupling to nucleon number $g_S^2$ of a yet-undiscovered scalar with mass $m_5$.  The shaded region is excluded.  Other limits shown are Pokotilovski 2006 \cite{pokotilovski2006constraints}, Mohideen et al. 1998 \cite{mohideen1998precision}, Haddock et al. 2018 \cite{haddock2018search}, and Kamiya et al. 2015 \cite{kamiya2015constraints}.}
    \label{fig:Results}
\end{figure*}

\begin{table*}[hbt!]
    \centering
    {\renewcommand{\arraystretch}{1.2}
    \begin{tabular}{|c|c|c|c|c|}
    \hline
      $hkl$ & $b(Q) / b(0) - 1$ &  Unc. Stat.   &  Unc. Sys.    & Unc. Total \\ \hline
     (111)  &-0.011$\,$075      & 0.000$\,$035  & 0.000$\,$038  & 0.000$\,$051 \\ \hline
     (220)  &-0.030$\,$203      & 0.000$\,$059  & 0.000$\,$036  & 0.000$\,$069 \\ \hline
     (400)  &-0.060$\,$596      & 0.000$\,$080  & 0.000$\,$049  & 0.000$\,$094 \\ \hline
    \end{tabular}
    }
    \linespread{2}
    \caption{\textbf{Measured structure factors and uncertainties (68~\% confidence intervals).}  All results are scaled to 295.5~K.}
\end{table*}

\clearpage


\setcounter{page}{1}

\begin{centering}
\vspace{2 cm}
\Huge{Supplementary Materials for} \\ 
\vspace{1 cm}
\Large{Pendell\"{o}sung Interferometry Probes the Neutron Charge Radius, Lattice Dynamics, and Fifth Forces} \\
\vspace{2 cm}
\normalsize{
Benjamin Heacock, Takuhiro Fujiie, Robert W. Huan, Albert Henins, \\
Katsuya Hirota, Takuya Hosobata, Michael G. Huber, Masaaki Kitaguchi, \\
Dmitry A. Pushin, Hirohiko Shimizu, Masahiro Takeda, Robert Valdillez, \\
Yutaka Yamagata, Albert R. Young\\
}
\vspace{0.5 cm}
\normalsize{Correspondence to:  benjamin.heacock@nist.gov.}

\end{centering}

\vspace{1cm}

\subsection*{This PDF file includes:}

Materials and Methods

\noindent
Supplementary Text

\noindent
Figs. S1 to S5

\noindent
Tables S1 to S2

\clearpage

\setcounter{equation}{0}
\renewcommand{\theequation}{S\arabic{equation}}

\section*{Materials and Methods}

\subsection*{Sample Preparation}

A diagram of the pendell\"{o}sung sample is shown in Fig.~\ref{fig:CrystalThick}A.  The neutron beam enters the main face of the crystal.  Aligning the surface normal of the major face to the $(\bar{2} 2 0)$ planes made available a number of reflections suitable for pendell\"{o}sung studies, where the lattice vector $\vec{H} = \vec{Q}$ is parallel to the surface.  For three of these reflections, flats were cut so that the sample could fit between the blades of the interferometer.  The beam profiles for the interferometry and pendell\"{o}sung measurements are also shown in Fig.~\ref{fig:CrystalThick}A.  The (111), (220), and (004) reflections were all measured.  Note that throughout the text, the (004) reflection in Fig.~\ref{fig:CrystalThick}A is referred to as the (400) reflection.

The pendell\"{o}sung sample was aligned to the Bragg planes and cut from a float zone grown silicon ingot.  The alignment of the $(\bar{2}20)$ Bragg planes to the major surface is $ < 0.25$~degrees.  The major surfaces of the sample were finished using the ultra-high precision grinding capabilities at the RIKEN Center for Advanced Photonics.  The grinding was performed in a similar manner to the interferometer described in \cite{heacock2019measurement}.  Approximately 5~$\mu$m of material was chemically etched from each side of the crystal using a 60:1 ratio mixture of nitric and hyrdrofluoric acids to remove machining damage.  Two-dimensional profiles of the change in crystal thickness were measured for each side of the crystal using a helium-neon laser interferometer.  The two profile measurements were added together to form a map of the crystal's thickness variation, which is shown in Fig.~\ref{fig:CrystalThick}C. The profile measurements were completed before the crystal was etched; however, while etching changes the absolute thickness of the sample, it is unlikely that the profile changed appreciably.

\subsection*{Pendell\"{o}sung Interferograms}
\label{sec:pendinf}

Pendell\"{o}sung interferograms were measured for the (111), (220), and (400) reflections by rotating the crystal about the $\vec{H}$-axis ($\theta_P$ in Fig.~1C) in discrete steps and recording count rates in a $^3$He detector.  Interferograms were measured with the crystal rotated by $\pm \theta_B$.  The resulting data was assigned statistical uncertainties from Poisson counting statistics and globally fit to

\begin{equation}
    \mathcal{I}(\theta_P) = A + B \cos \left [2 \pi (C_0 \pm \Delta C_0) \left ({1 \over \cos (\theta_P - \theta_{P0})} -1   \right ) + 2 \pi( C \pm \Delta C) + \phi_\mathrm{calc}   \right ],
    \label{eqn:PfitFinal}
\end{equation}

\noindent
for each reflection, where $\theta_P$ is the tilt of the crystal read out from the positing stage (Fig.~1C); $C_0$ is fixed to the nearest integer fringe number; $\phi_\mathrm{calc}$ is a fixed, geometry-dependent phase factor that is discussed below; $C$, $\Delta C_0$, and $\Delta C$ are shared fit parameters between the two interferograms; $A$, $B$, and $\theta_{P0}$ are fit parameters unique to each interferogram; and $\pm$ refers to the $\pm \theta_B$ interferograms.  In this procedure, $C$ is the average pendell\"{o}sung fringe number

\begin{equation}
C = \left ( \frac{2 g_{hkl} \tan{\theta_B}}{\pi a^2 \sqrt{h^2 + l^2 + k^2}} \right) D b(Q)
\label{eqn:Cp1}
\end{equation}
 
\noindent 
and the parameter of interest, where $g_{hkl} = \{ 8, \; \sqrt{32}, \; \mathrm{or} \; 0 \}$ corresponds to even, odd, and forbidden reflections, respectively.  The resulting statistical uncertainties for the pendell\"{o}sung phase shift $\phi_P = 2 \pi C $ for each reflection are reported in Table~S1.  Systematic uncertainties associated with the pendell\"{o}sung interferograms are discussed below and also reported in Table~S1 in order of appearance. 

Performing the global fits for the $\pm \theta_B$ interferograms removes systematic uncertainties associated with misalignments between the rotation axis of the positioning stage and $\vec{H}$ (see ``Crystal Manipulation" in the Supplementary Text below). To compute $D b(Q)$, the measured pendell\"{o}sung fringe number is divided by the term in parentheses in Eqn.~\ref{eqn:Cp1}, with $\theta_B$ taken by the angular position of the rotation stage.  The stage's embedded angular encoder had a resolution of $7.5 \times 10^{-5}$~degrees and accuracy of $ \pm 4.2 \times 10^{-4}$~degrees which we took as the 68~\% confidence interval.

 An array of slits was required to suitably prepare the beam and isolate the pendell\"{o}sung signal as shown in Fig.~\ref{fig:schem}.  Widths of $S_0 = 1.4$~mm, $S_1 = 1.1$~mm and $S_2 = 1.6$~mm were selected with the distances between the slits $L_1 = 1.42$~m and $L_2 = 0.48$~m.  The phase offset of the pendell\"{o}sung interferogram $\phi_\mathrm{calc}$ is a function of slit size and position.  The computed phase shifts were 0.78~rad, 0.84~rad, and 0.79~rad for the (111), (220), and (400) reflections, respectively.  The correction from the limiting case of an infinitely-wide slit (see ``Slit Size" in the Supplementary Text below) $\phi_\mathrm{calc} = {\pi \over 4} \simeq 0.785$ was smaller than the experimental uncertainties for all the reflections. 

The $S_0$ and $S_1$ slits remained stationary, while the $S_2$ slit and detector for the forward-diffracted beam could be translated into position.  The central value for the translation of $S_2$ was measured by translating the slit through the direct beam, and finding the centroid of the resulting peak.  The slit was moved into the forward diffracted beam by translating the slit by $ \pm D \sin \theta_B$.  Lead screw and stepper motor accuracy of the translation stage causes a $4$~$\mu$m systematic uncertainty in the slit position, and the angular misalignment of the translation stage is taken to be at most 2.5~degrees, which causes a $ < 8$~$\mu$m misalignment of $S_2$.  Given the slope of $\phi_\mathrm{calc}$ versus position of $S_2$, the systematic uncertainty caused by the positioning error of the final slit is much smaller than the total experimental uncertainty.

The height of the pendell\"{o}sung crystal was set by placing a cadmium slit on the crystal mount and lifting the mount through the beam.  The crystal was aligned in the horizontal direction separately for both Bragg angles and each reflection.  This was accomplished by translating the crystal through the beam past its edge while monitoring the diffracted beam.  When the neutron beam misses the crystal, only the background intensity is measured.  As the crystal edge is translated into the beam, the diffracted beam intensity is approximately linearly increasing.  Diffracted beam counts as a function of translation were measured, and the linear portion of the curve was fit to a line.  The intersection of this line with the background count rate was determined to be the edge of the crystal relative to the neutron beam.  Using the known beam width and crystal dimensions, the crystal was then translated from its edge to the target position, which was chosen to correspond to the location of the crystal measured by the neutron interferometer in the forward scattering measurement.

Interferograms were measured with the crystal translated by $\pm 1$~mm along $\vec{H}$ from the target value. There was no measurable difference in the pendell\"{o}sung phase shifts as a function of translation.  Depending on the local parallelism of the crystal, a systematic uncertainty for misalignment of the beam is considered,  but this is ascribed to the forward-scattering phase shift.  A clerical error caused some of the pendell\"{o}sung interferograms for the $+ \theta_B$ (111) reflection to be shifted by 5~mm away from the target position (Fig.~\ref{fig:CrystalThick}).  To compensate for the translation, the average thickness difference between the $\pm \theta_B$ interferogram positions was computed by averaging the crystal profile map over the beam areas outlined in Fig.~\ref{fig:CrystalThick}.  This was repeated with the beam profiles pseudo-randomly co-translated according to a two dimensional normal distribution (horizontal and vertical translations) with $\sigma_\mathrm{trans}=1$~mm.  The average thickness difference for the two beam profiles was 0.30(26)~$\mu$m, where the average and uncertainty is the average and standard deviation of the resulting thickness difference distribution. The difference between the thickness of the pendell\"{o}sung beam profiles and the thickness of the crystal at the target position gave a total correction of 0.09(8)~$\mu$m, resulting in a $9(8) \times 10^{-6}$ relative correction and uncertainty applied to the total pendell\"{o}sung phase shift for the (111) reflection.  There was no statistically-significant difference between the impacted and non-impacted (111) data sets. 

The temperature of the samples was controlled with a heater plate suspended above the crystal.  For the (220) and (400) measurements, a second heater was placed next to the crystal opposite from the larger heater.  Both heaters were controlled in a PID feedback loop using temperature probes.  Two other calibrated temperature probes were placed on the crystal a known distance apart, typically about 15~mm.  The temperature gradient was taken from the difference in measured temperature for the two probes.  The phase of the interferogram was corrected according to \cite{hart1966pendellosung}, where for small gradients the pendell\"{o}sung fringe number is altered by $C \rightarrow C(1+p_T^2/6)$ where

\begin{equation}
    p_T = \left (  \frac{ \pi a \, D (h^2 + l^2 + k^2) \alpha_T}{4 \, b(Q) \, g_{hkl} \sin \theta_B \cos \theta_B} \right )\frac{\partial T}{\partial x}
    \label{eqn:pT}
\end{equation}

\noindent
for the coefficient of thermal expansion of silicon $\alpha_T = 2.6 \times 10^{-6} \; \mathrm{K}^{-1}$ and temperature gradient in the direction perpendicular to the Bragg planes $\partial T / \partial x$.  All corrections were smaller than the statistical uncertainties and given a systematic uncertainty according to the standard deviation of the measured temperature gradients.
 
All results were scaled to 295.5~K using the BvK model predicted slope of $B$ at room temperature $dB / dT = 0.0014$~\AA${}^{2} \, \mathrm{K}^{-1}$ at 295.5~K. The sensitivity to $d B / dT$ was very slight (of order 0.001~\AA$^2 \, \mathrm{K}^{-1}$), because all interferograms were taken at similar temperatures. Specifically, $T_{111} = 295.78(20)$, $T_{220} = 295.67(8)$, and $T_{400} = 295.30(8)$, where the uncertainties are taken as the standard deviation of temperature measurements taken during the pendell\"{o}sung interferograms for each reflection.  The lattice constant is also a function of temperature; however, silicon's coefficient of thermal expansion is small enough for this to be a negligible effect.  A value of $a = 5.431020$~\AA \cite{kessler2017lattice} was used when computing $D b(Q)$.

To assess whether a second Bragg condition (called a ``parasitic" reflection) being excited as a function of crystal tilt could bias the pendell\"{o}sung interferograms, the Bragg conditions for all of the available Bragg vectors were plotted for 2.2~\AA~and 4.4~\AA~as a function of $\theta_B$ and $\theta_P$.  The (220) and (400) pendell\"{o}sung interferograms were unaffected.  Fig.~\ref{fig:111Pats} shows this plot for the (111) reflection.  A number of parasitic reflections cross the pendell\"{o}sung-interfering reflection between $\theta_P = \pm 2.5$~degrees for the (111) reflection measured at 4.4~\AA.  The additional crossings at $\pm 10$~degrees do not effect the interferograms, which were limited in range to $\pm 8$~degrees.  Note that the lines in Fig.~\ref{fig:111Pats} are for a monochromatic beam and do not account for the blurring that occurs due to the spread of the incoming beam's momentum distribution.

Because all of the parasitic reflections that cross the 4.4~\AA~(111) reflection are from 2.2~\AA~neutrons, the shorter wavelength neutrons were eliminated for a subset of the (111) data by placing a beryllium filter upstream.  Interferograms taken with and without the 2.2~\AA~component present agree within statistics - except in the portion of the interferogram affected by parasitic reflections.  Both (111) data sets were used in the result, but the data set without the beryllium filter was masked in the affected region (Fig.~\ref{fig:111Infgms}).

\subsection*{Forward Scattering Measurement}

A forward scattering interferogram is generating by rotating a fused silica phase flag that intersects both beam paths.  The phase flag is rotated through a small span in angle ($\pm 1.5$~degrees), and the difference in optical path length through the phase flag for the two interferometer beam paths is linear in phase flag rotation.  This is a standard technique in neutron interferometry \cite{rauch2015neutron}.  The result is a sinusoidal signal which may be fit to

\begin{equation}
    \mathcal{I} = A + B \cos \left [ C_I \delta  + \phi \right ]
    \label{eqn:InfFit}
\end{equation}

\noindent
where $A$, $B$, $C_I$, and $\phi$ are fit parameters, and $\delta$ is the angle of phase flag rotation.  The fitted phase of the interferogram is composed of two parts

\begin{equation}
    \phi = \phi_0(t) + \phi_I ,
\end{equation}

\noindent
where $\phi_0(t)$ is the intrinsic phase of the interferometer and can drift as a function of time and $\phi_I$ is the sample phase shift of interest.  To isolate the phase shift of the sample, interferograms must be measured with the sample both in and out of the interferometer beam paths. The difference in fitted phases of the interferograms is then due to the sample.   Typical interferograms for the sample in and out of the interferometer are shown in Fig.~1D.  Like the pendell\"{o}sung interferograms, the goodness of fit for the forward scattering interferograms is well-described using Poisson counting statistics for each data point.  The associated statistical uncertainty is given in Table~S2 along with the systematic uncertainties listed in order of appearance in the following discussion.  

The interferometer phase shift

\begin{equation}
    \phi_I = \Bigg ( N_\mathrm{Si} \lambda \Bigg ) b(0)D = \left ( {16 \over a^2 \sqrt{3}}\sin \theta_B \right ) b(0) D
    \label{eqn:phiI}
\end{equation}

\noindent
is proportional to $\sin \theta_B$, where $N_\mathrm{Si}$ is the number density of silicon, and $\theta_B$ is the Bragg angle of the \textit{interferometer}.  The $\sqrt{3} = \sqrt{h^2 + l^2 + k^2}$ in the denominator comes from the fact that the interferometer uses the (111) Bragg reflection.  By measuring interferograms with the interferometer turned to both $\pm \theta_B$ and averaging the measured phase shifts, the linear terms in $\delta \theta_B$ again average out, and the quadratic terms are small enough to be neglected.  The $b(0) D$ factor is isolated using the known silicon lattice constant $a$ and $\theta_B$ taken from the positioning stage, which contributes a $7 \times 10^{-6}$ systematic relative uncertainty to the sample phase shift.  The $\theta_B$ axis for the interferometer was aligned to be perpendicular to $\vec{H}$ of the interferometer by measuring the angular locations of the $\pm \theta_B$ Bragg peaks as a function of tilt about the $\theta_T$ axis (perpendicular to the major faces of the blades) of the interferometer.  The result is a parabola for $2 \theta_B $ as measured by the $\pm \theta_B$ Bragg peak centroids versus $\theta_T$.  The interferometer is brought to the center of this parabola with an estimated uncertainty of 0.04~degrees for $\theta_T$, which contributes a negligible uncertainty to $\theta_B$ for the $\pm \theta_B$ interferograms.  This is a common alignment technique for Bragg-diffracting crystals \cite{rauch2015neutron}.

To account for the temporal drifts of $\phi_0(t)$ a series of sample In-Out-Out-In interferograms are taken.  Linear drifts are canceled by this scheme.  Additionally, a reduced $\chi^2$ may be computed for each In-Out-Out-In measurement

\begin{equation}
    \chi^2_\mathrm{red} = \sum_{i=1}^4 {(\phi_i - \bar{\phi} \pm \Delta \bar{\phi} )^2 \over 2 \sigma_{\phi_i}^2 }
\end{equation}

\noindent
where $i$ indexes the interferogram; $\sigma_\phi$ is the uncertainty from fitting the interferogram to Eqn.~\ref{eqn:InfFit}; $\bar{\phi}$ is the average fitted phase of the four measurements; $\Delta \bar{\phi} = (\bar{\phi}_\mathrm{out}-\bar{\phi}_\mathrm{in})/2$ is the average phase difference; the $\pm$ sign depends on whether $i$ corresponds to the sample in or out of the interferometer; and the factor of 2 in the denominator accounts for the two degrees of freedom in each In-Out-Out-In sequence.  To account for rapid, higher-order drifts in $\phi_0(t)$, the uncertainty of the the phase shift for each In-Out-Out-In sequence was expanded by $\sqrt{\chi^2_\mathrm{red}}$ if $\chi^2_\mathrm{red}$ was greater than unity.  The error expansion could be as large as 4.5, as shown in Fig.~\ref{fig:InfPhaseTime}.  Note that the increased uncertainty in the weighted average of the phase shift is absorbed into the statistical uncertainty in Table~S2.

In addition to the temporal phase associated with $\phi_0(t)$, there can be a thermal link between the sample and the interferometer.  This creates a systematic deviation in the measured sample phase shift and has been the subject of recent study at the NIOF \cite{haun2020precision}.  To estimate the size of this effect, In-Out-Out-In interferogram sequences were measured with no sample in the crystal holder.  No material enters the neutron beam, and the phase shift associated with this measurement is due only to thermal effects.  Initially, the thermal phase was measured to be about 6~degrees; however, a thin layer of aluminum placed between the sample and the interferometer was found to disrupt the thermal link enough for the thermal phase to be less than one degree.  The thermal phase was measured in the $+ \theta_B$ geometry before measuring the sample phase shifts and again after all the sample phase shifts were measured a few days later, this time in the $- \theta_B$ geometry.  The resulting thermal phase measurement was $0.6 \pm 0.9$~degrees, and each forward scattering phase shift was independently corrected by this value.  The uncertainty was treated as uncorrelated, because the thermal phase can also drift as a function of time, albeit much more slowly than the intrinsic interferometer phase \cite{haun2020precision}.

All of the forward scattering interferograms were measured at $294 \; $K. The lattice constant was taken to be $a = 5.431017$~\AA~when computing $D b(0)$ from the sample phase shift, with $a$ adjusted to $294 \,$K.  When taking the ratio of $b(Q) D$ and $b(0) D$, the result was multiplied by a factor of $[1 + \alpha \Delta T ] = (1 - 3.9 \times 10^{-6} ) $ where $ \Delta T = T_I - T_P$ is the difference in sample temperature for the forward scattering and pendell\"{o}sung portions of the experiment.  The relative shift of $3.9 \times 10^{-6}$ was much smaller than the uncertainties for $b(Q) / b(0)$.

The measured phase shift was corrected by the phase shift of the atmosphere displaced by the sample \cite{ioffe1998precision,rauch2015neutron}

\begin{equation}
    \phi_\mathrm{air} = N_\mathrm{air} b_\mathrm{air} \lambda D.
\end{equation}

\noindent
At $\phi_\mathrm{air} = 105.3(8)$~degrees, this correction is not small.  The uncertainty of the correction is from considering a combination of temperature, barometric pressure, and relative humidity.  The uncertainty contributed by that of the scattering lengths of each gas species is negligible compared to the environmental effects.

The thickness appearing in Eqn.~\ref{eqn:phiI} is the optical thickness.  If the crystal slab is misaligned from the beam by some angle $\epsilon$, then the optical thickness is $D / \cos \epsilon \simeq D (1 + {1 \over 2} \epsilon^2 ) $.  The quadratic dependence of $\epsilon$ is utilized to align the sample to the interferometer beam.  Before doing the longer set of In-Out-Out-In measurements, the phase shift of the sample is measured as a function of sample tilt (horizontal axis perpendicular to the beam) and rotation (vertical axis).  The sample tilt and rotation were set to the central values of the resulting sample phase shift parabolas.  The uncertainties for $\epsilon$ were on the order of 0.1~degrees; the resulting relative change in the sample phase shift of $\sim 1.5 \times 10^{-6}$ is negligible.  This is another standard practice in neutron interferometry \cite{ioffe1998precision, rauch2015neutron}.

The height of the pendell\"{o}sung sample relative to the beam inside of the interferometer was set by tipping off the crystal holder for the forward scattering measurement to the crystal while it was still in the holder for the pendell\"{o}sung measurement.  The horizontal translation of the sample was set by translating a cadmium slit in the interferometer beam path for both $\pm \theta_B$.  The translational alignment was only performed in the (111) geometry.  The alignment to the the (220) and (400) sample volumes was accomplished using a set of tapped holes in the plate to which the crystal holder was secured; the sample could be removed and re-fixed to the mount in a known relative alignment.  The absolute translational accuracy of this method is estimated to be about 0.9~mm, given the machining tolerances of the mount. The resulting systematic uncertainty associated with the interferometer and pendell\"{o}sung beam irradiating slightly different crystal volumes was computed using the local thickness gradient from moving the beam masks in Fig.~\ref{fig:CrystalThick}C.  The change in thickness uncertainties were 0.23~$\mu$m, 0.21~$\mu$m, and 0.35~$\mu$m for the (111), (220), and (400) sections of the crystal, respectively. The resulting systematic uncertainties in the phase shift were the largest of the experiment and similar in size to the overall statistical uncertainties.

The phase shifts were used to compute a crystal thickness using the scattering length of silicon \cite{ioffe1998precision} for each portion of the crystal.  These values all fell within a 2~$\mu$m window and were found to be consistent $(p=0.14)$ with the thickness variation map as shown in Fig.~S1C.  While the interferometer beam profiles are wider than the relevant pendell\"{o}sung region, the central portion of the interferometer beam has the highest interference fringe visibility.  Additionally, the forward scattering phase shift was measured with the sample translated over a 2~mm to 4~mm range for all three of the sample sections, and no phase gradients with respect to sample translation were resolvable.

\subsection*{Fitting Procedure}

The $\chi^2$ sum of weighted residuals was minimized with respect to $B$ and $\langle r_n^2 \rangle$ when fitting $b(Q)/b(0)$ to Eqn.~1.  An additional term was added to the $\chi^2$ to account for a slight anharmonic correction to the Debye-Waller factor, where $e^{-W} \rightarrow e^{-W}[1 \pm T_a (h k l)]$ for odd structure factors \cite{dawson1967anharmonic}.  The anharmonic contribution $T_a(hkl)$ depends on temperature and is proportional to the product $h \times k \times l$.  The $\pm$ corresponds to whether $h + k + l \pm 1$ is divisible by four.  We use a relative correction of $ \pm 3.1 \times 10^{-5} (h \times k \times l)$, as was measured using the (222) forbidden neutron reflection from silicon at room temperature \cite{roberto1974diffraction}.  While this does not agree with data taken at higher temperatures (650~K to 1650~K) and extrapolated to room temperature \cite{roberto1974diffraction,hastings1975high,flensburg1999lattice} which would predict $T_a(hkl) = 1.5 \times 10^{-5} (h \times k \times l)$, we use the room temperature data, because the classical anharmonic model that was used breaks down with decreasing temperature as quantum zero-point vibrations become relevant.  Nevertheless, we expand the uncertainty for $T_a$ from $0.3 \times 10^{-5}$ to $1.0 \times 10^{-5}$ in the $\chi^2$ to account for the discrepancy.  The anharmonic correction only impacts the (111) data at the $1.5 \times 10^{-5}$ level, which is to be compared to the total relative uncertainty of $5 \times 10^{-5}$.

In addition to the anharmonic term, an additional weighted residual for $\langle r_n^2 \rangle$ or $B$ was added to the $\chi^2$ minimization when constraining the model for comparison between literature $B$ or $\langle r_n^2 \rangle $ values, respectively, as discussed in detail in the Supplementary Text and shown in Fig.~2.  To compute the constraints for $\alpha_G$, additional terms for both $\langle r_n^2 \rangle$ and $B$ were added to the $\chi^2$ with $\lambda_5$ fixed.  For $\langle r_n^2 \rangle$, the PDG value was used in keeping with other constraints on the exclusion plot (Fig.~3).  The value and weight of $B = 0.4714(57) \,$ \AA$^2$ was chosen such that the 68~\% confidence region encapsulates $B$ as computed from x-ray data, regardless of whether one assumes a single $B$ parameter for all electron shells or $B_\mathrm{1s}$.  This is a conservative estimate for $B$ which encompasses all values shown in Fig.~2.

\clearpage

\section*{Supplementary Text}

\subsection*{Comparison with Prior Determinations of $\langle r_n^2 \rangle$ and $B$}

The PDG suggested value for $\langle r_n^2 \rangle_\mathrm{PDG} = -0.1161(22) \, \mathrm{fm}^2$ \cite{PDG2018} uses results from epithermal neutron transmission in lead \cite{kopecky1997neutron, koester1995neutron} and bismuth \cite{kopecky1997neutron,koester1995neutron,aleksandrov1986neutron}, as well as thermal neutron scattering asymmetry in noble gasses \cite{krohn1973reconsideration}.  The PDG applies a 0.03~fm$^2$ correction to the noble gas scattering measurements, which was recommended by \cite{koester1976measurement} due to Schwinger scattering.  However, the origins and reasoning behind this correction are not clear, and the authors of the paper recommending the correction later utilize the uncorrected values when comparing their results to previous work \cite{koester1995neutron}.  Other authors that report compilations of $\langle r_n^2 \rangle$ also do not correct the noble gas values \cite{kopecky1997neutron}.  Schwinger scattering is a second order effect which does not typically impact thermal scattering measurements, whereas epithermal measurements must supply a small correction \cite{kopecky1997neutron}.  Additionally, the PDG uncertainty is scaled by a factor of 1.3, yet all of the tension lies in different measurements of bismuth, where the sum of the three weighted residuals is $\chi^2 = 6.5$ ($p=0.039$).  The lack of internal consistency could be due to differing approaches for the nuclear resonance or solid state corrections that are required for epithermal transmission measurements, which could uniquely impact each nuclide.  If the bismuth data is expanded by $\sqrt{\chi^2 / \mathrm{DOF}} = 1.8$ for two Degrees of Freedom DOF and the Schwinger scattering correction is omitted from the noble gas scattering data, then $ \langle r_n^2 \rangle _\mathrm{avg} = -0.1137(13)$~fm$^2$, with $\chi^2 = 4.7$, ($p=0.86$) for the seven measurements (we separate the noble gas scattering measurements into their four separate values).

If $B$ is allowed to vary freely, our determination of $\langle r_n^2 \rangle = -0.1101(89) \, \mathrm{fm}^2$ is consistent with both $\langle r_n^2 \rangle_\mathrm{avg}$ and $\langle r_n^2 \rangle_\mathrm{PDG}$.  However, constraining $B$ to neutron inelastic scattering data fit to a BvK model (scaled to 295.5~K) $B_\mathrm{BvK} = 0.4725(17) \, \mathrm{\AA}^2$ \cite{flensburg1999lattice} yields a value of $\langle r_n^2\rangle = -0.1011(66) \, \mathrm{fm}^2$ which no longer agrees with $\langle r_n^2 \rangle_\mathrm{avg}$, but does agree with $\langle r_n^2 \rangle _\mathrm{EFT}$. However, it was noted by the authors of the $B_\mathrm{BvK}$ value that assigning weights to the BvK fit was not straightforward \cite{flensburg1999lattice}.  The uncertainty given to $B_\mathrm{BvK}$ depends on expanding the relative uncertainties of the measured phonon frequencies to be no less than 0.5~\% and then additionally expanding the uncertainty by a factor of $\sqrt{\chi^2 / \mathrm{DOF}} = 2.3$.  The validity of expanding uncertainties in this way assumes that the errors of the individual data points are underestimated and that any unaccounted sources of uncertainty are random.  It is considerably more likely that either (1) the BvK model is invalid at the level of precision of the experimental data or (2) an unknown systematic error in the experimental data shifted the reported phonon frequencies relative to the true values, in which case the prescription for expanding experimental uncertainties could miss a systematic deviation in $B$.  
This is further motivated by recent work done with neutron inelastic scattering in silicon which produces a density of states which has different qualitative features then that of the BvK model.  Compare, for example, Fig.~1 of~\cite{hellman2013temperature} which uses a fully anharmonic phonon model to Fig.~2 of \cite{flensburg1999lattice}.  The tension ($p=0.018$) between $B_\mathrm{BvK}$ and our result $B = 0.4767(6)$~\AA$^2$ when $b(Q)/b(0)$ is constrained according to $\langle r_n^2 \rangle_\mathrm{avg}$, thus suggests that the BvK model may be insufficient for describing thermal vibrations in silicon.  We eagerly anticipate new recommended values for $B$ from neutron inelastic scattering data fit to more sophisticated anharmonic phonon models so that we may compare our results to such theories.

Values of $B$ derived from fitting x-ray data to measured structure factors vary widely, depending on which data sets are used and what assumptions are made about the static electron structure factors.  Some analyses also allow differing $B$ for each atomic electron shell \cite{deutsch1989thermal}.  We sample two such approaches in Fig.~2.  First, we use the theoretical static structure factors from \cite{erba2013accurate}, whose ab initio computed $B$ (scaled to 295.5~K) is also reported in Fig.~2.  Then we fit to compiled x-ray structure factor data \cite{cummings1988redetermination} assuming (1) the same Debye-Waller factor for all electron shells and (2) a separate Debye-Waller factor for the 1s shell and the rest of the electrons.  The 1s shell static structure factor was taken from \cite{deutsch1989thermal}, which also performed an analysis with multiple $B$.  However, the authors set four different $B$, one for the 1s, 2s, 2p, and bonding electrons, arriving at values of $B = \{0.531(3),0.941(45),0.283(12),0.58(15)\}$~\AA$^2$ for the 1s, 2s, 2p, and bonding electrons, respectively.  We elected to use only two $B$ for the following reasons:  The static structure factors in \cite{deutsch1989thermal} empirically account for bonding distortions of the atomic orbitals, but apply the entire correction to the valence electrons.  Some of this correction should also be applied to the 2s and 2p electrons which hybridize into 2sp$^3$ bonding and antibonding molecular orbitals.  For this reason, separate $B$ should not be applied to the 2s and 2p electrons.  The 2sp$^3$ orbitals are oriented relative to the bonding directions, whereas the 1s electron shell retains spherical symmetry.  Because silicon's primitive cell contains two atoms of opposite bonding parities, optical phonons distort the bonding and anti-bonding molecular orbitals, causing a breakdown in the rigid atom approximation which affects the valence and 2sp$^3$ electrons differently than the 1s electrons.  Ideally, separate $B$ would be applied to the 1s, 2sp$^3$, and valence electrons.  However, because we do not have access to static structure factors partitioned in this way (which as pointed out by \cite{erba2013accurate} is a somewhat arbitrary partitioning of the total electron density), we separate $B$ for the 1s and remaining electrons.  This is a reasonable estimation, as the contribution of the valence electrons to the total static electron structure factors is small ($<1$~\%) for all structure factors except (111), such that the fitted $B$ for the 2sp$^3$ and valence electrons combined should be very close to that of only the 2sp$^3$ electrons.  This is why the uncertainty for $B$ for the bonding electrons from \cite{deutsch1989thermal} has a large uncertainty.  Finally, instead of computing bonding distortions empirically, we use the static structure factors from sophisticated lattice calculations \cite{erba2013accurate}, which were not available when researchers began considering separate $B$ for each electron shell.

Our determination using a single $B_\mathrm{xray} = 0.4664(7)$~\AA$^2$ is consistent with that of other x-ray analyses \cite{cummings1988redetermination}, which is considerably lower than the both $B_\mathrm{BvK}$ value and theoretical calculations \cite{erba2013accurate}.  The $B$ computed for just the 1s electrons is consistent with both $B_\mathrm{BvK}$ and this work, albeit with uncertainties that are much larger than when only a single $B$ parameter is considered.  The large variation between $B$ computed for each electron shell in \cite{deutsch1989thermal} seems unlikely, as it implies that the 2s electrons have a mean square displacement that is over three times larger than the 2p electrons.  Partitioning according the principle quantum number $n$ creates a more subtle result with $B_\mathrm{1s} = 0.4743(28)$~\AA$^2$ and $B_{n \geq 2} = 0.4604(21)$~\AA$^2$, where the slight difference between the two presumably corresponds to bonding distortions from high-frequency optical modes, where the reduction of the mean square displacement for $n \geq 2$ is logical. Repeating our measurements with germanium will provide an interesting additional tests of separate $B$ as thermal bonding distortions will presumably be altered due to hybridization of the bonding electrons with 3d orbitals.

\subsection*{Slit Size}

Pendell\"{o}sung oscillations occur as a function of wavelength and crystal thickness but also as a function of distance from the center of the forward-diffracted beam.  There are two limiting cases, a narrow slit and a wide slit case, with the two cases shifted by $\pi \over 4$ in the pendell\"{o}sung phase shift.  The narrow and wide slit cases are defined by 

\begin{equation}
S_1 \; \mathrm{and} \; S_2 \ll \sin \theta_B \sqrt{\Delta_H D}  \; \; \; \;  \mathrm{narrow} 
\end{equation}
\begin{equation}
S_1 \; \mathrm{or} \; S_2 > \sin \theta_B \sqrt{\Delta_H D} \; \; \; \;  \mathrm{wide}
\label{eqn:Cases}
\end{equation}

\noindent
with $S_1$ and $S_2$ the widths of the slits immediately before and after the crystal, respectively, as defined in Fig.~\ref{fig:schem}.  The pendell\"{o}sung length $\Delta_H \sim 50 \, \mu$m is the spatial period of oscillation as a function of crystal thickness $C = D / \Delta_H$.  The previous high-precision pendell\"{o}sung measurements employed the narrow slit case \cite{shull1972spherical}, whereas we use the wide slit case to maximize intensity.  For intermediate slit sizes, $\phi_\mathrm{calc}$ in Eqn.~\ref{eqn:PfitFinal} shifts smoothly by from ${\pi \over 2}$ for the narrow slit case to  ${\pi \over 4}$ for the wide slit case.  The exact value of $\phi_\mathrm{calc}$ can be computed using numerical integration, and small corrections to the limiting cases are sometimes required both here and for previous measurements \cite{shull1968observation}.

\subsection*{Crystal Manipulation}
\label{sec:xtalMan}

Upon tilting the crystal about $\vec{H}$, the intensity in the diffracted beam shows oscillations of the form

\begin{equation}
    \mathcal{I}(\theta_P) = A + B \cos \left ( {2 \pi C \over \cos (\theta_P - \theta_{P0}) } + \phi_\mathrm{calc}  \right ) 
    \label{eqn:pendInf}
\end{equation}

\noindent
where $A$, $B$, $C$, and $\theta_{P0}$ are fit parameters.    In order to generate a pendell\"{o}sung interferogram, the reciprocal lattice vector $\vec{H}$ must be aligned to tilt axis of the positioning stage.  To perform this alignment, the crystal is fixed to a goniometer with three rotation axes.  A Bragg peak occurs as the crystal is rotated by $\theta_B$ in Fig.~1C through the Bragg condition where  $\sin \theta_B =  H \lambda /( 4 \pi)$. If $\vec{H}$ and the pendell\"{o}sung tilt axis are aligned, then rotating by some angle $\theta_P$ about $\vec{H}$ will not change the angular position of the Bragg peak.  If the tilt axis and $\vec{H}$ are misaligned vertically, then the angle of elevation of $\vec{H}$ above the diffraction plane $\epsilon$ takes on a nonzero value.  The projection of $\vec{H}$ in the plane perpendicular to rotation axis of the $\theta_B$ positioning stage is then dependent on $\theta_P$.  The result is that the angular position of the Bragg peak depends linearly on $\theta_P$, where the peak centroid changes more rapidly for larger magnitude $\epsilon$.

To align $\vec{H}$ of the crystal in its mount to the tilt stage that controls $\theta_P$, a three dimensional alignment scan was used.  First the crystal was tilted by $\theta_T$ about the axis perpendicular to the major face (Fig.~1C) to intentionally induce a nonzero $\epsilon$.  Bragg peak centroids were then measured for a few values of $\theta_P$.  The Bragg peak centroids versus $\theta_P$ were fit to a line.  The slope of this line for each value of $\theta_T$ also forms a line, and the value of $\theta_T$ for which the Bragg peak centroid is unaltered by a change in $\theta_P$ can be computed from the fits.  Using this alignment method $\epsilon$ was limited to less than 0.1~degrees.

A nonzero $\epsilon$ impacts the pendell\"{o}sung interferogram by causing $\theta_P$ to take on a wavelength dependence.  The pendell\"{o}sung fringe number is proportional to $\tan \theta_B$.  Letting $\theta_B$ be dependent on $\theta_P$, and also expanding $1 / \cos \theta_P$ to second order in $\theta_P$ the fitted pendell\"{o}sung fringe number $C$ deviates from the true value of $D / \Delta_H$

\begin{equation}
     { C(\epsilon) \over \cos \theta_P} = \frac{D}{\Delta_H} \left ( 1 + {1 \over 2} \theta_P^2 + {\epsilon \theta_P \over \cos \theta_B \sin \theta_B }  + \mathcal{O}(\epsilon^2,\theta_P^3) \right ) .
    \label{eqn:thetaPexp}
\end{equation}

\noindent
This has a minimum with respect to $\theta_P$ at

\begin{equation}
    \theta_{P0} = - {\epsilon \over \cos \theta_B \sin \theta_B }.
    \label{eqn:thetaP0}
\end{equation}

\noindent
The net result is that the interferogram are shifted left or right in Fig.~1E.  If $\epsilon$ is nonzero, then the interferogram shifts in the opposite direction for $\pm \theta_B$.  Because pendell\"{o}sung interferograms are taken at both $ \pm \theta_B$ as a way of measuring the incoming neutron wavelength (see Fig.~1A), the difference in the fitted center of the interferograms $\theta_{P0}$ gives an extra indication of $\epsilon$.  In addition to shifting the curve left or right, the fitted pendell\"{o}sung fringe number is altered by

\begin{equation}
    C(\epsilon) = \frac{D}{\Delta_H} \left [  1 - {1 \over 2} \theta_{P0}^2 + \mathcal{O} (\theta_{P0}^3)\right ]  =  \frac{D }{ \Delta_H} \left [ 1 - {1 \over 2} \left ({ \epsilon \over \cos \theta_B \sin \theta_B }  \right )^2   + \mathcal{O} (\epsilon^3) \right ]
\end{equation}

\noindent
Fortunately, with the previously described alignment of $\theta_T$ to minimize $\epsilon$, this correction is very small at ${1 \over 2 } \theta_{P0}^2 < 5 \times 10^{-6}$.

The alignment of $\vec{H}$ to the $\theta_P$ tilt axis in the diffraction plane ($\theta_B$ axis in Fig.~1C) depends on the machining tolerances of the parts holding the crystal and is estimated to be $ \delta < 2.5 \; \mathrm{deg}$.  The fitted pendell\"{o}sung fringe number as a function of $\theta_P$ has a slight $\delta$-dependence

\begin{equation}
   { C(\delta) \over \cos{\theta_P} } = { D / \Delta_H \over \sin^2{\delta} + \cos^2{\delta} \cos{\theta_P}} \left ( 1 + { (1 - \cos \theta_P) \sin \delta  \over \sin \theta_B \cos \theta_B}  \right ) .
    \label{eqn:Cchng}
\end{equation}

\noindent
The change in the $\cos \theta_P$ term in the denominator is a geometric effect (as $\delta$ increases, the thickness change as a function of $\theta_P$ decreases), and the $(1- \cos \theta_P)$ term is due to a change in wavelength.  The wavelength term is linear in $\delta$ and therefore dominates.  Expanding Eqn.~\ref{eqn:Cchng} to second order in $\theta_P$

\begin{equation}
    {C(\delta) \over \cos \theta_P} = \frac{D}{\Delta_H} \left [  1 + {1 \over 2}  \left ( 1 + {\sin \delta \over \sin \theta_B \cos \theta_B}    \right ) \theta_P^2 + \mathcal{O}(\theta_P^4) \right ],
\end{equation}

\noindent
it is clear that the relationship between the pendell\"{o}sung fringe position $C$ and the rate at which the phase is modulated by changing the crystal tilt $\theta_P$ is altered by the $\delta$ angular misalignment.  To account for this, the pendell\"{o}sung fringe position and $ 1 / \cos \theta_P$ terms in Eqn.~\ref{eqn:pendInf} were allowed to be different

\begin{equation}
    \mathcal{I} ( \theta_P )  = A +  B \cos \left [ 2 \pi C_0 \left ( {1 \over \cos \theta_P} -1  \right ) + 2 \pi C  + \phi_\mathrm{calc}  \right ],
    \label{eqn:pendfitalt}
\end{equation}

\noindent
which prevents a nonzero $\delta$ misalignment from biasing the fitted pendell\"{o}sung fringe position $C$.  While the extra parameter $C_0$ increases the uncertainty in the fitted pendell\"{o}sung fringe position, the $\delta /( \sin \theta_B \cos \theta_B )$ term carries opposite signs for the $ \pm \theta_B$ interferograms.  A combined fit for the pendell\"{o}sung interferograms at $\pm \theta_B$ may be constrained such that $C \rightarrow C \pm \Delta C$ and $C_0 \rightarrow C_0 \pm \Delta C_0$, resulting in a global fitting function Eqn.~\ref{eqn:PfitFinal}, where the fitted $C$ is the average of the $\pm \theta_B$ interferograms and unbiased by $\delta$ or the difference in wavelength $\delta \lambda$ from the $\pm \theta_B$ scans.  The average phase modulation parameter $C_0$ was constrained to be the nearest integer pendell\"{o}sung fringe number, where changing $C_0$ by $\pm 1$ fringe was found to have a negligible effect on the fitted $C$.



\clearpage

\renewcommand{\thefigure}{S1}

\begin{figure}
    \centering
    \includegraphics[width=0.6 \textwidth]{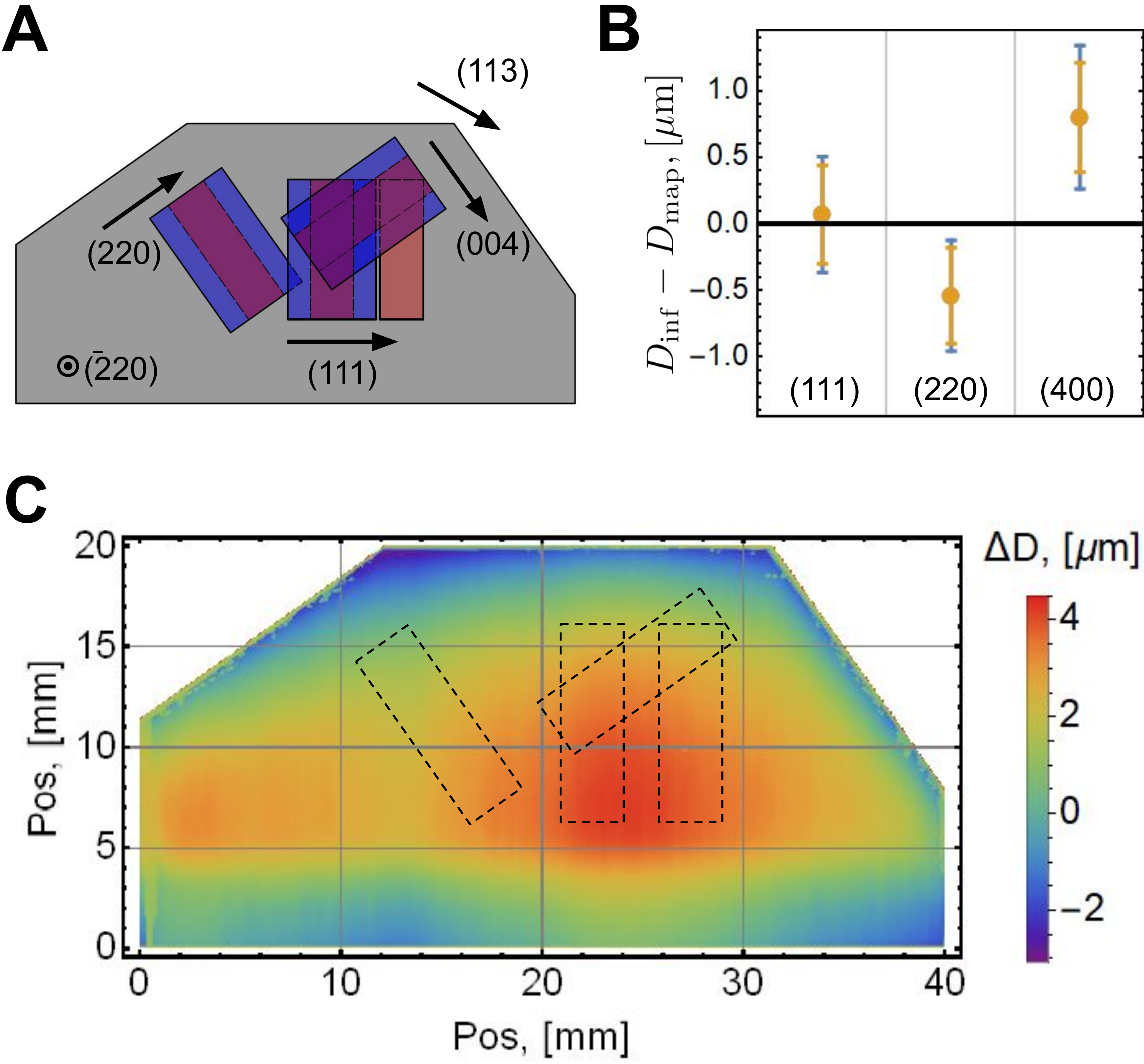}
    \linespread{2}
    \caption{\textbf{Diagram of the pendell\"{o}sung crystal.} \textbf{(A)} The irradiated portions and Bragg plan orientation of the pendell\"{o}sung crystal. The blue sections are the approximate interferometer beam profile, which is most sensitive to the crystal thickness at its center, corresponding to the red sections which are the areas probed by the pendell\"{o}sung measurements. \textbf{(B)} Difference in thickness measured using the interferometer and the thickness variation map.  Larger error bars are for the total forward scattering error budget, and the smaller error bars do not include the systematic uncertainty associated with translational alignment of the sample. \textbf{(C)} The thickness variation map was measured before etching using a helium-neon laser.  The outlines of the pendell\"{o}sung beam profiles are shown as dashed lines.   A portion of the (111) data was shifted by 5~mm to the right, as shown.  A small correction was applied based on the profile map.  See text for details.}
    \label{fig:CrystalThick}
\end{figure}

\renewcommand{\thefigure}{S2}

\begin{figure}
\centering
\includegraphics[width=0.5\textwidth]{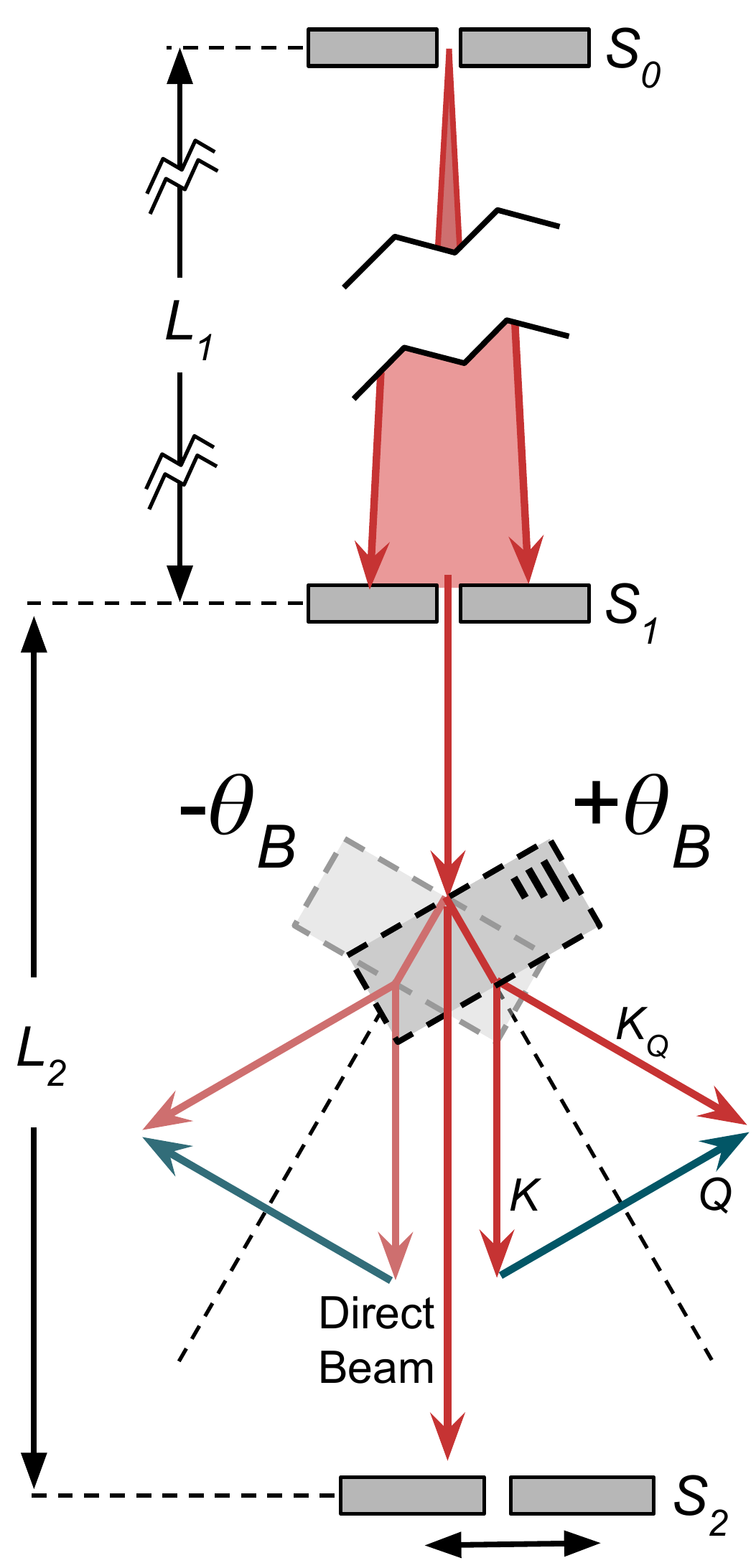}
    \linespread{2}
    \caption{
    \textbf{Slit geometry for measuring pendell\"{o}sung interferograms.} Slit widths of $S_0 = 1.4$~mm, $S_1 = 1.1$~mm and $S_2 = 1.6$~mm and separations of $L_1 = 1.42$~m and $L_2 = 0.48$~m were utilized.
    }
    \label{fig:schem}
\end{figure}

\renewcommand{\thefigure}{S3}

\begin{figure}
    \centering
    \includegraphics[width=0.9\textwidth]{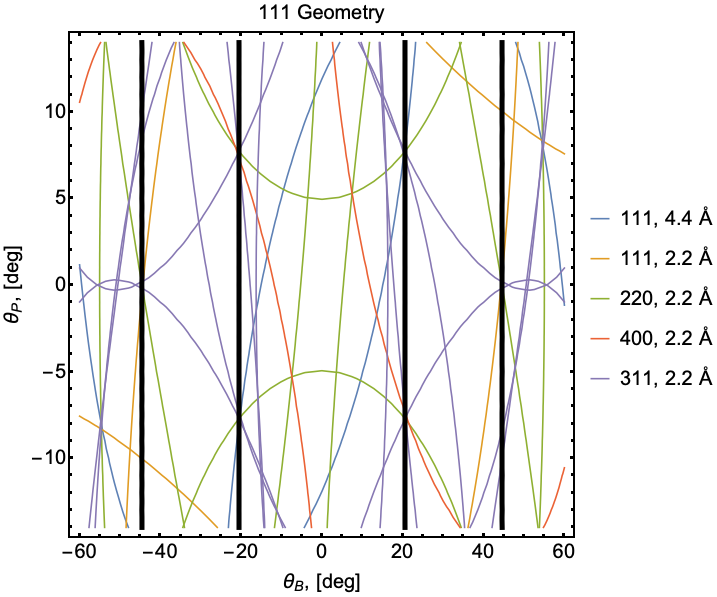}
    \linespread{2}
    \caption{\textbf{ \boldmath Contours of $ \theta_P $ and $\theta_B$ where an available Bragg condition of the pendell\"{o}sung crystal is satisfied for 2.2~\AA~or 4.4~\AA~neutrons in the (111) experimental geometry. \unboldmath}  The black lines at $\theta_B \sim \pm 20$~degrees and $\theta_B \sim \pm 45$~degrees denote where (111) pendell\"{o}sung fringes may be measured for the  2.2~\AA~and 4.4~\AA~components of the beam, respectively.  Inserting an upstream beryllium filter removes the parasitic reflections that affect the 4.4~\AA~pendell\"{o}sung interferograms.}
    \label{fig:111Pats}
\end{figure}

\renewcommand{\thefigure}{S4}

\begin{figure}
    \centering
    \includegraphics[width=0.8\textwidth]{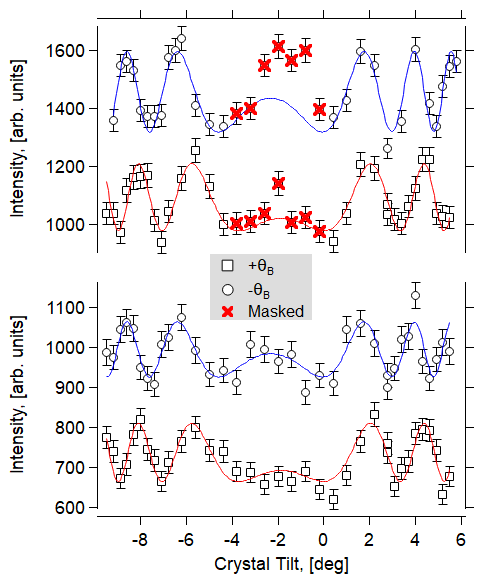}
    \linespread{2}
    \caption{\textbf{ \boldmath Summed pendell\"{o}sung interferograms for the (111) reflection using $\lambda = 4.4$~\AA~neutrons. \unboldmath}  The top pair of curves is with no beryllium filter.  The bottom pair is with the beryllium filter in place.  The systematic effect of parasitic reflections is clear.  A shift in the baseline of some curves was added for clarity.}
    \label{fig:111Infgms}
\end{figure}

\renewcommand{\thefigure}{S5}

\begin{figure}
    \centering
    \includegraphics[width = \textwidth]{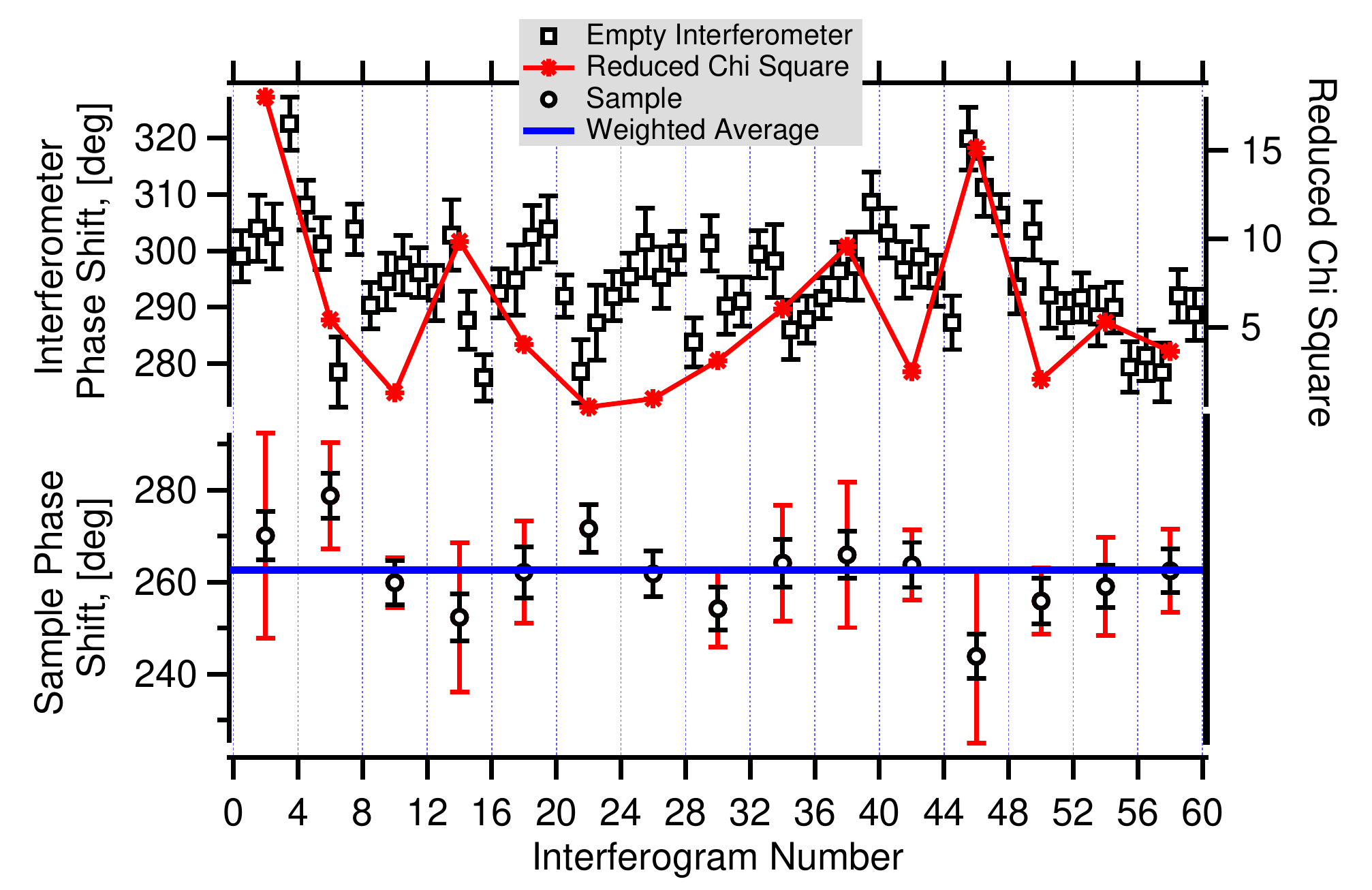}
    \linespread{2}
    \caption{\textbf{Typical interferometer phase and sample phase shifts versus time.}  Additionally, the reduced $\chi^2$ of each In-Out-Out-In measurement is plotted.  The red error bars for the sample phase shift are expanded according the the reduced $\chi^2$.  See text for details.}
    \label{fig:InfPhaseTime}
\end{figure}

\clearpage

\renewcommand{\thetable}{S1}

\begin{table}
    \centering
    
    {\renewcommand{\arraystretch}{1.2}
    \begin{tabular}{l|l |c|c|c}
        Source          & Uncertainty   & $\Delta\phi_P$ (degrees) & $\sigma_P / \phi_P \times 10^{-5}$  & $hkl$ \\ \hline \hline
                             &               &       & $1.5 $  & 111 \\
    Bragg Stage, $\theta_B$  & $4.3 \times 10^{-4}$~degrees &  & $ 1.6 $    & 220 \\
                        &               &      & $1.5 $    & 400 \\ \hline
                        &               & 0.0 & $ 0.5$ & 111 \\
    $S_2$ Translation   & 10~$\mu$m     & 3.1 & $ 0.6$ & 220 \\
                        &               & 0.3 & $ 0.6$ & 400 \\ \hline
    Crystal Profile     & 0.08~$\mu$m  & 0.9 & 0.8 & 111 \\ \hline
                        & $0.1 \times 10^{-5}$ & 0.2 & $0.1$  & 111 \\
    Temperature Gradient, $p^2_T / 6$   & $0.2 \times 10^{-5}$ & 0.4 & $0.2 $  & 220 \\
                        & $ 2.1 \times 10^{-5}$ & 3.7 & $ 2.1 $    & 400 \\ \hline
                        & 0.20~K   &  -1.0  & $0.6$  & 111 \\
    Absolute Temperature   & 0.06~K &  -1.0 & $0.6 $  & 220 \\
                        &  0.08~K &  3.2 & $ 1.5 $    & 400 \\ \hline \hline
                        &               &       & $1.9$  & 111 \\
    \textbf{Total Sys.}          &               &       & $1.8$  & 220 \\
                        &               &       & $3.0$  & 400 \\ \hline \hline
                        & 1.9~degrees     &       & $2.0 $  & 111 \\
    \textbf{$\phi_P$, Stat.}  & 3.3~degrees     &       & $5.4$  & 220 \\
                        & 6.3~degrees      &       & $7.8$  & 400 \\ \hline \hline \hline
                        &               &       & \textbf{2.7}  & 111 \\
    \textbf{Total}         &               &       & \textbf{5.7}  & 220 \\
                        &               &       & \textbf{8.4}  & 400 \\ \hline \hline \hline
    \end{tabular}
    }
    
    \linespread{2}
    \caption{\textbf{Uncertainty budget for the pendell\"{o}sung interferogram measurements.}  The correction to the measured pendell\"{o}sung phase shift is $\Delta \phi_P$.  The total phase shifts for $\phi_P$ were approximately $10 \times 10^4$~degrees, $6 \times 10^4$~degrees and $8 \times 10^4$~degrees for the (111), (220), and (400) reflections, respectively.}
    \label{tab:pendunc}
    
\end{table}

\renewcommand{\thetable}{S2}

\begin{table}
    \centering
    
    {\renewcommand{\arraystretch}{1.2}
    \begin{tabular}{l|l |c|c|c}
        Source          & Uncertainty   & $\Delta \phi_I$ (degrees) & $\sigma_I / \phi_I \times 10^{5}$  & $hkl$ \\ \hline \hline
    Bragg Stage, $\theta_B$  & $4.3 \times 10^{-4}$~degrees &  & $ 0.7 $    & All \\ \hline
    Thermal Phase        & 0.9~degrees      & 0.6 & $1.7$ & All \\ \hline
    Air Scattering, $\phi_\mathrm{air}$ & 0.8~degrees       & 105.3 &  $1.6$ & All \\ \hline
                        &               &   & $ 2.3$ & 111 \\
    Crystal Translation   & 0.9~mm        &   & $ 2.1$ & 220 \\
                        &               &   & $ 3.6$ & 400 \\ \hline \hline
                        &               &       & $3.4 $  & 111 \\
    \textbf{Total Sys.}          &               &       & $3.2$  & 220 \\
                        &               &       & $4.3$  & 400 \\ \hline \hline
                        & 1.5~degrees     &       & $2.9 $  & 111 \\
    \textbf{$\phi_I$, Stat.}  & 1.5~degrees     &       & $2.9$  & 220 \\
                        & 1.8~degrees      &       & $3.5$  & 400 \\ \hline \hline \hline
                        &               &       & \textbf{4.4}  & 111 \\
    \textbf{Total}         &               &       & \textbf{4.3}  & 220 \\
                        &               &       & \textbf{5.5}  & 400 \\ \hline \hline \hline
    \end{tabular}
    }

\linespread{2}
\caption{\textbf{Uncertainty budget for the forward scattering measurements.}  The overall phase shift of the sample is approximately $5.1 \times 10^{4}$~degrees.}
    
\label{tab:infunc}
\end{table}


\end{document}